\documentclass[superscriptaddress,
showpacs,preprintnumbers,nofootinbib,eqsecnum,
amsmath,amssymb,aps,11pt, groupedaddress]{revtex4-2}
\usepackage{amsmath}
\usepackage{graphics}
\usepackage{epstopdf}
\usepackage{subfigure}
\usepackage{braket}
\usepackage{bm}
\usepackage{dcolumn}
\usepackage{cancel}
\usepackage[dvipdfmx,final]{graphicx}
\input{colordvi.tex}
\usepackage{longtable}
\usepackage{tabularx}
\usepackage{cleveref}
\usepackage{mathtools}
\usepackage[utf8]{inputenc}
\usepackage{xcolor}
\usepackage{soul}
\usepackage{pifont}

\usepackage{tabularx}
\usepackage{multirow, makecell, cellspace, bigstrut}
\usepackage{pict2e}
\usepackage{booktabs}
\usepackage{diagbox}

\newcolumntype{C}{>{\centering\arraybackslash}X}

\newcommand{\s}{\\ \vspace*{-3.5mm}}

\newcommand{\rb}[2]{\raisebox{#1}[-#1]{#2}}

\begin{document}

\begin{flushright}
KIAS-Q23023
\end{flushright}

\title{\large \bf Hunting for Hypercharge Anapole Dark Matter
        in All Spin Scenarios}

\author{Seong Youl Choi$^1$}
\email{sychoi@jbnu.ac.kr}
\author{Jaehoon Jeong$^2$}
\email{jeong229@kias.re.kr}
\author{Dong Woo Kang$^1$}
\email{dongwookang@jbnu.ac.kr}
\author{Seodong Shin$^1$}
\email{sshin@jbnu.ac.kr}
\affiliation{
$^1$Laboratory for Symmetry and Structure of the Universe,
Department of Physics, Jeonbuk National University, Jeonju,
Jeonbuk 54896, Korea \\
$^2$ School of Physics, Korea Institute for Advanced Study,
Seoul 02455, Korea
}

\begin{abstract}
\vskip 0.5cm
\noindent
We conduct a combined analysis to investigate dark matter (DM)
with hypercharge anapole moments, focusing on scenarios where
Majorana DM particles with spin 1/2, 1, 3/2, and 2 interact exclusively with Standard Model particles through U(1)$_{Y}$
hypercharge anapole terms for the first time.
For completeness, we construct general effective U(1) gauge-invariant three-point vertices.
These enable the generation of hypercharge gauge-invariant interaction vertices for
both a virtual photon $\gamma$ and a virtual $Z$ boson with
two identical massive Majorana particles of any non-zero
spin $s$, after the spontaneous breaking of electroweak
gauge symmetry. For complementarity, 
we adopt effective operators tailored to each dark matter spin allowing crossing
symmetry. We calculate the relic abundance, analyze current
constraints and future sensitivities from dark matter
direct detection and collider experiments, and apply
the conceptual naive perturbativity bound. Our estimations
based on a generalized vertex calculation
demonstrate that the scenario with a higher-spin
DM is more stringently constrained than a lower-spin DM,
primarily due to the reduced annihilation cross-section and/or
the enhanced rate of LHC mono-jet events. As a remarkable
outcome, the spin-2 anapole DM scenario is almost entirely
excluded, while the high-luminosity LHC exhibits high
sensitivities in probing spin-1 and 3/2 scenarios, except for
a tiny parameter range of DM mass around 1 TeV.
A significant portion of the remaining parameter space in the
spin-1/2 DM scenario can be explored
through upcoming Xenon experiments, with more than 20 ton-year
exposure equivalent to approximately 5 years of running
the XENONnT experiment.

\end{abstract}

\maketitle

\newpage

\section{Introduction}
\label{sec:introduction}

The nature of dark matter (DM) remains one of the greatest
puzzles in particle physics and cosmology, accounting for
approximately a quarter of the total energy content in the
Universe. Exploring the non-gravitational interactions of
DM with Standard Model (SM) particles offers a direct path
to uncovering new structures and symmetries.
This exploration can be pursued through three distinct
experimental approaches: direct detection, indirect detection,
and collider experiments. These three different types of
experiments can be complementary to one another, and hence it
is extremely helpful to combine the results and apply them to
a single effective operator for DM-SM interactions. This is
a reasonable and widely applicable approach in scenarios where
the corresponding effective field theory (EFT) is
valid~\cite{Beltran:2010ww, Cao:2009uw, Fox:2011pm,
Belyaev:2018pqr}.

Certainly, for a proper EFT description, it is required that
the energy scale of all processes under consideration be well
below the masses of the mediating particles, and that their
interactions respect the established low-energy (global and/or
gauge) symmetries of the SM~\cite{Glashow:1961tr,
Weinberg:1967tq, Salam:1968rm, Fritzsch:1973pi}.
As the TeV energy scale, which is higher than the electroweak
(EW) scale, is currently being probed and the SM with its
SU(3)$_C \times$ SU(2)$_L \times$ U(1)$_Y$ gauge symmetry has
been firmly established, particularly with the discovery of
the Higgs boson~\cite{Aad:2012tfa, Chatrchyan:2012ufa},
it is appropriate to maintain both the hypercharge U(1)$_Y$
symmetry and the other non-Abelian gauge symmetries,
SU(3)$_C$ and SU(2)$_L$. For instance, the importance of
considering the hypercharge U(1)$_Y$ rather than the
electromagnetic U(1)$_{\rm EM}$ as a valid U(1) gauge symmetry
has been clearly and persuasively demonstrated from
various physics perspectives in recent
work~\cite{Arina:2020mxo}.

The non-gravitational interaction between DM and SM particles
through an electromagnetic (EM) form factor, for DM with
nonzero spin, is realized through a higher-dimensional
operator. Consequently, this provides an interesting test bed
for the EFT approach in DM studies. Additionally, the EM form
factor induces unexpected (and suppressed) EM interactions
of DM with SM particles, without the DM being directly
charged. The phenomenological effects of such interactions
were first discussed in Ref.~\cite{Pospelov:2000bq}.

In this paper, we adopt an EFT approach for scenarios
involving CPT self-conjugate Majorana DM with spin 1/2 to 2,
where interactions with SM particles are exclusively mediated
through U(1)$_Y$ hypercharge anapole terms. For neutralino DM
in supersymmetric models, which has been the most preferred
and  theoretically well-motivated DM candidate over
the past decades, the corresponding anapole term is
the only allowed U(1) form factor, and hence it is
worthwhile to study in
detail~\cite{Cabral-Rosetti:2014cpa, Cabral-Rosetti:2015cxa,
Ibarra:2022nzm}.
The first Kaluza-Klein excitation of the hypercharge gauge
boson is a typical candidate for a spin-1 Majorana particle
interacting with the SM particles through hypercharge
anapole terms. The scenario involving spin-1 DM particles
has been investigated in various
works~\cite{Cheng:2002ej, Servant:2002aq, Hubisz:2004ft,
Birkedal:2006fz, Hambye:2008bq, Hisano:2010yh,
Davoudiasl:2013jma, Gross:2015cwa, Karam:2015jta,
Flacke:2017xsv, Choi:2019zeb, Elahi:2019jeo, Abe:2020mph,
Nugaev:2020zcv, Elahi:2020urr}.
The spin-3/2 DM has been well studied as gravitino in
supergravity, and also in various effective
theories~\cite{Ellis:1983ew,Khlopov:1984pf,Ellis:1984eq,
Olive:1984bi,Yu:2011by,Ding:2012sm,Savvidy:2012qa,Ding:2013nvx,
Khojali:2016pvu,Khojali:2017tuv,Chang:2017dvm,Savvidy:2012qa,
Garcia:2020hyo,Wu:2022gni,Goyal:2022cmz,Kaneta:2023uwi}.
The massive graviton has been widely discussed as a spin-2 DM
candidate in extra dimensional models and bigravity
theories~\cite{Arkani-Hamed:1998sfv,Feng:2003nr,Dubovsky:2004ud,
Pshirkov:2008nr,Aoki:2016zgp,Babichev:2016hir,Babichev:2016bxi,
Aoki:2017cnz,Aoki:2017ffl,Chu:2017msm,GonzalezAlbornoz:2017gbh,
Garny:2017kha,Aoki:2017ixz,Cai:2021nmk,Wu:2022gni,
Manita:2022tkl,Gorji:2023cmz}.

Previously, several aspects of EM anapole DM interaction terms
have been studied in the context of relic abundance
measurements as well as in searches at direct detection,
indirect detection, and collider experiments for the spin-1/2
case~\cite{Ho:2012bg, Ho:2012br, Gao:2013vfa,
Geytenbeek:2016nfg, Latimer:2017lwm, Alves:2017uls,
Kang:2018oej, Florez:2019tqr, Bose:2023yll} and the spin-1
case~\cite{Hisano:2020qkq, Chu:2023zbo}.
A comprehensive analysis of the U(1)$_Y$ anapole DM
scenario with a spin-1/2 Majorana particle was recently
conducted~\cite{Arina:2020mxo}. On the contrary, the
analysis for the spin-1 case was restricted so far to the EM
anapole DM scenario, as the focus was only on the direct
DM detection capability. 
In this context, it is worthwhile to perform a generic 
analysis accommodating and characterizing 
the hypercharge anapole DM particle of any nonzero spin
with combined experimental probes all together.

In our general analysis, we construct effective
U(1)$_Y$ gauge-invariant three-point vertices of a hypercharge
gauge boson $B$ and two identical on-shell Majorana particles
with any non-zero spin.\footnote{For clarity, the term `
Majorana particle' is conventionally used to represent
a spin-1/2 self-conjugate particle; however, we will
generalize it to include self-conjugate particles of any spin
without loss of generality, as done in the
work~\cite{Boudjema:1990st}.}
These vertices accommodate an arbitrary spin $s$ and non-zero
mass $m$, generating interaction vertices with not only
a virtual photon $\gamma$ but also a virtual gauge boson $Z$
after electroweak symmetry breaking (EWSB). A concise overview
of the distinction between our analysis targets and those of
other studies is shown in Table~\ref{tab:summary_of_works_on_anapole_dm} of
our summary and conclusion section.

Having outlined the general three-point vertices,
our in-depth numerical analysis is specifically targeted
towards four scenarios where the DM particle spin is set
to be 1/2, 1, 3/2, and 2, while qualitatively exploring the
implications for the scenarios with its spin larger than 2.
We incorporate the relic abundance value determined by the
Planck collaboration~\cite{Planck:2018vyg}, the up-to-date
result from the DM direct detection experiment XENONnT
with approximately 1.1 ton-year
exposure~\cite{XENON:2023cxc}, and the LHC experiments with the
integrated luminosity of 139 fb$^{-1}$~\cite{Shiltsev:2019rfl,
ATLAS:2021kxv,CMS:2021far}, as well as the so-called naive
perturbativity bound (NPB) making our EFT approach
valid.\footnote{Conceptually, there could also be unitarity
bounds on the couplings for each annihilation mode, but these
are quantitatively much weaker than the naive perturbativity
limits.} In addition, we estimate the projected
sensitivities of the high luminosity LHC (HL-LHC) experiment
with the full run of 3 ab$^{-1}$ integrated
luminosity~\cite{Assmann:2023dwx} and those of the future
XENONnT with the 20 ton-year exposure.

This paper is organized as follows.
In Sec.~\ref{sec:anapole_vertices_a_majorana_particle}, we
derive the general effective hypercharge gauge-invariant
three-point vertices. These generate interaction vertices for
a virtual photon $\gamma$ and a massive gauge boson $Z$ with
two identical on-shell particles of any nonzero spin $s$
and mass $m$, following EWSB. The derivation is based on
an efficient and systematic algorithm for constructing the
covariant effective vertex for three particles of any spin
and mass~\cite{Choi:2021ewa, Choi:2021qsb, Choi:2021szj}.
In Sec.~\ref{sec:dm_relic_abundance}, we calculate the
annihilation cross sections of two Majorana particles into
kinematically allowed pairs of SM particles. We then determine
the constraints on the effective coupling strengths for the
spin-1/2, 1, 3/2, and 2 cases from the observed DM
relic abundance.
Section~\ref{sec:collider_searches} is devoted to determining
constraints on the coupling strengths from recent LHC and
upcoming HL-LHC
experiments~\cite{Shiltsev:2019rfl, Assmann:2023dwx}.
We particularly focus on how these
constraints depend on the spin of the anapole DM Majorana
particle.
In Sec.~\ref{sec:direct_searches}, we explore constraints from
the up-to-date results of the DM direct detection experiment
XENONnT~\cite{XENON:2023cxc} and the projected sensitivities
from its highly enhanced exposure, and study
the implications for the coupling strengths.
Based on all the experimental constraints and an additional
theoretical constraint from the naive perturbativity bound,
we present an overall combined picture of the current
constraints and sensitivities in Sec.~\ref{sec:combined_constraints}.
This comprehensive analysis places special emphasis on systematically
delineating the distinct characteristics that vary
depending on the spin values.
We summarize the key points of our results and conclude in
Sec.~\ref{sec:summary_conclusion}.
Furthermore, Appendix~\ref{appendix:anapole_vertex_any_spin} provides
a concise and systematic algorithmic description for
constructing all the general three-point anapole vertices
efficiently and Appendix~\ref{appendix:relic_calculation_strategy}
provides a detailed explanation of the
numerical calculation strategy employed to determine
the DM relic abundance of dark matter.

\setcounter{equation}{0}

\section{Anapole vertices for two Majorana particles of any spin}
\label{sec:anapole_vertices_a_majorana_particle}

In this section, we present an efficient and systematic
algorithm for constructing covariant three-point anapole
vertices for two identical Majorana particles of any spin.
This algorithm allows us to perform a complete and systematic
characterization of all the spin values of the hypercharge
anapole DM.  Subsequently, we delve deeper into the specific case of
covariant vertices,
focusing on the spin-1/2, 1, 3/2, and 2 scenarios.
This detailed analysis encompasses both analytic
and numerical investigations to provide comprehensive insights and
findings.

\subsection{Aanapole three-point vertices of Majorana
            particles of any spin}

A Majorana particle, by definition, is a CPT self-conjugate particle,
which means it remains unchanged under the combined operations of
charge conjugation (C), parity reversal (P), and time reversal (T).
These particles do not possess any static charge or multipole moments,
as all the terms in their interaction Hamiltonian are CPT-odd.

The only permissible U(1) gauge-invariant interaction vertices between
a U(1) gauge boson and two identical massive Majorana particles
of nonzero spin are known as anapole-type moments~\cite{Boudjema:1990st}.
It is worth noting that no massless CPT self-conjugate Majorana particle
can have any U(1) gauge-invariant couplings unless its spin is 1/2.

The effective anapole three-point $\chi\chi B$ vertex of two
identical massive Majorana particles $\chi$ of any spin and a U(1)
gauge boson $B$ is in general given by the gauge invariant
Lagrangian
\begin{eqnarray}
   \mathcal{L}_{\rm anapole}^{}
=  {\cal J}_\mu\, \partial_{\nu} B^{\mu\nu}\,,
\label{eq:general_anapole_lagrangian}
\end{eqnarray}
with the 4-vector current ${\cal J}_\mu$ comprised of two Majorana-particle
fields and the field-strength tensor
$B^{\mu\nu}=\partial^{\mu}B^{\nu}-\partial^{\nu}B^{\mu}$ of the U(1)
gauge boson $B$.

\begin{figure}[ht!]
\vskip 0.5cm
\centering
\includegraphics[scale=1.3]{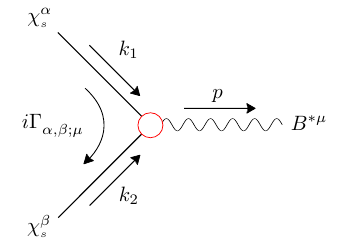}
\caption{\rm
   A diagram for the annihilation of two identical Majorana
   particles $\chi_{\tiny\mbox{$s$}}$ into an off-shell hypercharge gauge boson
   $B^*$. The combined momenta $p=k_1+k_2$ and
   $q=k_1-k_2$ are constructed by the combinations of incoming
   momenta $k_{1,2}$ of two Majorana particles.
   The $k_{1,2}$-dependent $\chi^\alpha_{\tiny\mbox{$s$}}$ and $\chi^\beta_{\tiny\mbox{$s$}}$
   are the wave functions of
   two identical Majorana particles with the particle symbol
   denoted by $\chi$ in the main text.
   The indices,
   $\alpha$ and $\beta$, stand collectively for the 4-vector
   indices, $\alpha=\alpha_1\cdots \alpha_n$ and
   $\beta=\beta_1\cdots \beta_n$, with $n=s-1/2$
   or $n=s$ for a half-integer or integer spin-$s$ Majorana
   particle. The curved arrow is for an arbitrary
   chosen fermion-number flow direction whose meaning is
   described in detail
   in Refs.~\cite{Denner:1992me,Denner:1992vza}.
}
\label{fig:diagram_xxB}
\end{figure}

Equation (\ref{eq:general_anapole_lagrangian}) enables us to
construct the effective conserved current for the annihilation
of two Majorana particles into a virtual vector boson depicted
in Fig.~\ref{fig:diagram_xxB}. Explicitly, the current can be
cast into the form
\begin{eqnarray}
  V_\mu (p,q)
= p^2 J_\mu(p,q) - p\cdot J(p,q)\, p_{\mu}\,,
\label{eq:conserved_vector_current}
\end{eqnarray}
with $p=k_1+k_2$ and $q=k_1-k_2$ in terms of the incoming
Majorana-particle momenta, $k_1$ and $k_2$, where the vector current
$J_\mu$ is nothing but the momentum-representation version
of the position-representation current ${\cal J}_\mu$ in
Eq.$\,$(\ref{eq:general_anapole_lagrangian}). Note that
the vector current $V_\mu$ automatically satisfies the U(1)
gauge invariance condition
$p^\mu V_\mu=0$. In the covariant formulation,
the vector current $V_\mu$ can be written as the products
between the wave tensors $\chi_{\tiny\mbox{$s$}}(k_1)$ and $\chi_{\tiny\mbox{$s$}}(k_2)$
of two Majorana particles and a covariant three-point
vertex $\Gamma$
\begin{eqnarray}
   V_\mu(p,q)
= \bar{\chi}^{\beta}_{\tiny\mbox{$s$}}(k_2)\,\Gamma_{\alpha,\beta;\mu}(p,q)\,
  \chi^{\alpha}_{\tiny\mbox{$s$}}(k_1)\,,
\label{eq:annihilation_current}
\end{eqnarray}
with respect to the arbitrarily chosen fermion-number-flow
arrow shown in Fig.$\,$\ref{fig:diagram_xxB},
where the indices, $\alpha$ and $\beta$, stand collectively
for the 4-vector indices, $\alpha=\alpha_1\cdots \alpha_n$
and $\beta=\beta_1\cdots \beta_n$, with $n=s-1/2$ or $n=s$
for a half-integer or integer spin-$s$ Majorana particle.
All the details about the wave
tensors~\cite{Behrends:1957rup,Auvil:1966eao,
Caudrey:1968vih,Scadron:1968zz,Chung:1997jn,Huang:2003ym}
and the general covariant vertex are included collectively in
Appendix~\ref{appendix:anapole_vertex_any_spin}.

As two identical Majorana particles annihilate into a virtual
gauge boson $B^*$, the covariant three-point vertex must
satisfy the so-called identical-particle (IP)
condition, (\ref{eq:fermionic_ip_relation})
or (\ref{eq:bosonic_ip_relation}), in the fermionic
or  bosonic case, respectively, as described in detail
in Appendix~\ref{appendix:anapole_vertex_any_spin}.
By employing the general covariant three-point vertices and
imposing the IP relation,
the anapole three-point vertex can be
cast into a compact square-bracket operator form:
\begin{align}
   [\Gamma_F]=&
\bigg(\frac{p^2}{\Lambda^2}\bigg)[A\,]
\sum_{\tau=0}^{n}
\bigg(\frac{p^2}{\Lambda^2}\bigg)^{\tau}
f^-_{\tau}[\,g\,]^{n-\tau}[S^0]^{\tau}
\nonumber\\
&+\bigg(\frac{p^2\sqrt{p^2}}{\Lambda^3}\bigg)
[V]
\sum_{\tau=1}^{n}
\bigg(\frac{p^2}{\Lambda^2}\bigg)^{\tau-1}
f^+_{\tau}[\,g\,]^{n-\tau}[S^0]^{\tau-1}
         \qquad\qquad\qquad\quad\;\;\;\;\,
         \mbox{ for fermions}\,,
\label{eq:general_fermionic_gamma}
\\
[\Gamma_B^{}]&=\sqrt{p^2}
\bigg(\frac{p^2}{\Lambda^2}\bigg)\,
\sum_{\tau=1}^{n}
\bigg(\frac{p^2}{\Lambda^2}\bigg)^{\tau-1}
\Big(
b^{-}_{\tau}[V^{-}]
+
b^{+}_{\tau}[V^{+}]
\Big)
[\,g\,]^{n-\tau}[S^0]^{\tau-1}\,
         \quad\;\;\,\mbox{ for bosons}\,,
\label{eq:general_bosonic_gamma}
\end{align}
in terms of $2s$ independent $f$ and $b$ couplings
for the spin-$s$ Majorana fermion with $s=n+1/2$ and boson with
$s=n$, respectively.
The cutoff scale $\Lambda$ is set forth explicitly to indicate
that the effective covariant vertex originates from a
higher-dimensional operator term in the given effective
Lagrangian or Hamiltonian.
Here, the square-bracket notations are introduced
for denoting the product of the basic helicity-related
operators as well as two derived operators in a compact
form with  all the four-vector and spinor index symbols
hidden.

Firstly, the two square-bracket operators, $[A]$ and $[V]$,
in Eq.$\,$\eqref{eq:general_fermionic_gamma} denote an
orthogonal axial-vector and vector currents, of which the
explicit forms are given by
\begin{align}
    &[A] \ \
\rightarrow \ \
   A_{\mu}\,\, =\,\,\gamma_{\bot \mu}\gamma_5,
   \label{eq:axial_vector_operator}
   \\
   &[V] \ \
\rightarrow \ \
   V_{\alpha\beta;\mu}\,\, =\,\,
   \hat{p}_{\alpha}g_{\bot \beta\mu}
   +\hat{p}_{\beta}g_{\bot \alpha\mu},
   \label{eq:vector_operator}
\end{align}
with an orthogonal gamma matrix
$\gamma_{\bot \mu}=g_{\bot \mu\nu}\gamma^{\nu}$
and an orthogonal metric tensor
$g_{\bot\mu\nu}=g_{\mu\nu}
-\hat{p}_{\mu}\hat{p}_{\nu}+\hat{q}_{\mu}\hat{q}_{\nu}$
involving two normalized momenta, $\hat{p}=p/\sqrt{p^2}$
and $\hat{q}=q/\sqrt{-q^2}$.
Secondly, due to the totally-symmetric property of
the wave tensors~\cite{Behrends:1957rup,Auvil:1966eao,
Caudrey:1968vih,Scadron:1968zz,Chung:1997jn,Huang:2003ym}
over all the four-vector indices, the $n$-th power products of the metric tensor $g$
and the basic scalar operator $S^0$ can be given in
a compact square-bracket form:
\begin{eqnarray}
  [\,g\,]^{n}
&\rightarrow &
  g_{\alpha_1,\beta_1}\cdots g_{\alpha_n\,\beta_n},
  \\[3pt]
  [S^0]^{n}
&\rightarrow &
   S^0_{\alpha_1,\beta_1}\cdots
   S^0_{\alpha_n\,\beta_n},
\label{eq:square_bracket_s_operators}
\end{eqnarray}
where the basic scalar operator $S^0$ is defined by
\begin{align}
   S^0_{\alpha\beta}  \,\,
&= \,\, \hat{p}_{\alpha }\hat{p}_{\beta},
\label{eq:basic_s_operator}
\end{align}
of which the repeated appearance at the vertices increases
the dimensions of the corresponding Lagrangians gradually.
It is compensated by introducing the proper power of
the cutoff scale $\Lambda$ along with the operator as shown
in Eqs.~\eqref{eq:general_fermionic_gamma}
and~\eqref{eq:general_bosonic_gamma}.
Thirdly, the other two derived basic
vector operators $[V^{\pm}]$ are defined by
\begin{align}
      [V^{\pm}] \ \
\rightarrow \ \
       V^\pm_{\alpha\beta;\mu} \,\,
= \,\, \hat{p}_{\beta}  S^{\pm}_{\alpha\mu}
      +\hat{p}_{\alpha} S^{\mp}_{\beta\mu},
\label{eq:derived_v_operator}
\end{align}
in terms of the normalized momentum $\hat{p}$ and the
basic scalar operators $S^\pm$ of which the explicit expression is
\begin{align}
 S^{\pm}_{\alpha\beta}=\frac12
 \big[g_{\bot \alpha\beta}\pm i\langle \alpha\beta\hat{p}\hat{q}\rangle],
\end{align}
with the angle-bracket
notation $\langle \alpha\beta \hat{p} \hat{q}\rangle=
\varepsilon_{\alpha\beta\rho\sigma}\hat{p}^{\rho}
\hat{q}^{\sigma}$ of a product between an anti-symmetric
Levi-Civita tensor and two normalized momenta $\hat{p}$ and
$\hat{q}$.

\subsection{Effective three-point anapole vertices for
the spin of 1/2, 1, 3/2 and 2}

Following the systematic derivation procedure for constructing
the general anapole vertices and using the general properties
of wave tensors described in
Appendix~\ref{appendix:anapole_vertex_any_spin}, we can
recast the covariant three-point anapole vertices extracted
from the general forms
in Eqs.~\eqref{eq:general_fermionic_gamma} and
\eqref{eq:general_bosonic_gamma}  effectively into the following form as
\begin{align}
    \Gamma^{[1/2]}_{\mu}
\,& \overset{\mbox{\tiny eff}}{=}\,
   \frac{p^2}{\Lambda^2} a_{1/2}\,\gamma_{\bot \mu}\gamma_5
   \qquad\qquad\qquad\qquad\qquad\qquad\qquad
\label{eq:spin-1/2_gamma}
\\[3pt]
   \Gamma^{[1]}_{\alpha,\beta;\mu}
\,& \overset{\mbox{\tiny eff}}{=}\,
   \frac{ip^2}{\Lambda^2}
   \Big[a_1\,
   \langle \alpha\beta\mu q \rangle_{\bot}
      - b_1\,
   (\,p_{\alpha}g_{\bot \beta\mu}+p_{\beta}g_{\bot \alpha\mu})
   \Big],
\label{eq:spin-1_gamma}
\end{align}
in terms of a single coupling $a_{1/2}$ in the spin-1/2
case and two independent couplings, $a_1$ and $b_1$,
in the spin-1 case with an orthogonal antisymmetric tensor
$\langle \alpha \beta \mu q\rangle_{\bot}
=g_{\bot\mu\nu}\, {\epsilon_{\alpha\beta}}^{\nu\sigma}\,
q_{\sigma}$, and
\begin{align}
\Gamma^{[3/2]}_{\alpha,\beta;\mu}
\,& \overset{\mbox{\tiny eff}}{=}\,
   \frac{p^2}{\Lambda^2} a_{3/2}\,\gamma_{\bot \mu}\gamma_5
   \,g_{\alpha\beta},
\label{eq:spin-3/2_gamma}
\\[3pt]
   \Gamma^{[2]}_{\alpha_1\alpha_2,\beta_1\beta_2;\mu}
\,& \overset{\mbox{\tiny eff}}{=}\,
   \frac{ip^2}{\Lambda^2}
   \Big[a_2\,
   \langle \alpha_1\beta_1\mu q \rangle_{\bot}
      - b_2\,
   (\,p_{\alpha_1}g_{\bot \beta_1\mu}+p_{\beta_1}g_{\bot \alpha_1\mu})
   \Big]\,g_{\alpha_2\beta_2},
\label{eq:spin-2_gamma}
\end{align}
in terms of a single coupling $a_{3/2}$ in the spin-3/2
case and two independent couplings, $a_2$ and $b_2$,
in the spin-2 case up to the leading order in
$1/\Lambda^2$. All the couplings  are in general
complex and the $a_i$ terms are parity-odd while the $b_j$
terms are parity-even where $i=1/2,1,3/2,2$ and $j=1,2$.
Although our numerical analysis is
confined up to spin 2, a simple extrapolation suggests
that a single coupling exists in any half-integer spin scenario,
while in the case of non-zero integer spin, there are two distinct
and independent couplings at the leading order of $1/\Lambda^2$,
as evident with Eqs.$\,$\eqref{eq:general_fermionic_gamma} and
\eqref{eq:general_bosonic_gamma}.

The effective U(1) gauge-invariant spin-1/2 and spin-1
anapole Lagrangians
corresponding to the vertices in Eqs.~\eqref{eq:spin-1/2_gamma}
and \eqref{eq:spin-1_gamma} can be re-constructed
by replacing each momentum with its corresponding derivative as
\begin{align}
   \mathcal{L}_{1/2}
\,=&\, \frac{a_{1/2}}{2\Lambda^2}
\,\bar{\chi}_{\tiny\mbox{$\frac12$}}\gamma^{\mu}\gamma_5
\chi_{\tiny\mbox{$\frac12$}}\,
         \partial_{\nu}B^{\mu\nu},
\label{eq:anapole_lagrangian-1/2}
\\[3pt]
   \mathcal{L}_{1}
\,=&\, \bigg[\frac{a_1}{2\Lambda^2} \epsilon_{\alpha\beta\mu\rho}
    \big[\chi_{\tiny\mbox{$1$}}^{\alpha}
        (\partial^{\rho}\chi_{\tiny\mbox{$1$}}^{\beta})
       -(\partial^{\rho}\chi_{\tiny\mbox{$1$}}^{\alpha})
         \chi_{\tiny\mbox{$1$}}^{\beta}\big]
       +\frac{b_1}{2\Lambda^2}
        \partial^{\rho}(
        \chi_{\tiny\mbox{$1$}\rho}
        \chi_{\tiny\mbox{$1$}\mu}
       +\chi_{\tiny\mbox{$1$}\mu}
        \chi_{\tiny\mbox{$1$}\rho}) \bigg]
         \partial_{\nu}B^{\mu\nu},
\label{eq:anapole_lagrangian-1}
\end{align}
in terms of the spin-1/2 and spin-1 Majorana fields, $\chi_{\tiny\mbox{$\frac12$}}$ and
$\chi_{\tiny\mbox{$1$}\mu}$, respectively.
Likewise, the U(1) gauge-invariant
spin-3/2 and spin-2
anapole Lagrangians
corresponding to the vertices in Eqs.~\eqref{eq:spin-3/2_gamma}
and \eqref{eq:spin-2_gamma}
are given as:
\begin{align}
\mathcal{L}_{3/2}
\,=&\, \frac{a_{3/2}}{2\Lambda^2}
\,\bar{\chi}_{\tiny\mbox{$\frac32$}\rho}\gamma^{\mu}\gamma_5
\chi_{\tiny\mbox{$\frac32$}}^{\rho}\,
         \partial_{\nu}B^{\mu\nu},
\label{eq:anapole_lagrangian-3/2}
\\[3pt]
   \mathcal{L}_{2}
\,=&\, \bigg[\frac{a_2}{2\Lambda^2} \epsilon_{\alpha\beta\mu\rho}
    \big[\chi^{\alpha\sigma}_{\tiny\mbox{$2$}}(\partial^{\rho}\chi^{\beta}_{\tiny\mbox{$2$}\,\sigma})
       -(\partial^{\rho}\chi^{\alpha\sigma}_{\tiny\mbox{$2$}})
       \chi^{\beta}_{\tiny\mbox{$2$}\,\sigma}\big]
       +\frac{b_2}{2\Lambda^2}
        \partial^{\rho}(
        \chi_{\tiny\mbox{$2$}\rho}^{\;\;\;\,\sigma}
        \chi_{\tiny\mbox{$2$}\mu\sigma}
       +\chi_{\tiny\mbox{$2$}\mu}^{\;\;\;\,\sigma}
        \chi_{\tiny\mbox{$2$}\rho\sigma}) \bigg]
         \partial_{\nu}B^{\mu\nu},
\label{eq:anapole_lagrangian-2}
\end{align}
in terms of the spin-3/2 and spin-2 Majorana fields, $\chi_{\tiny\mbox{$\frac32$}\mu}$ and
$\chi_{\tiny\mbox{$2$}\mu\nu}$, respectively.

In the following, we specify the U(1)
gauge boson to be the hypercharge U(1)$_Y$ gauge boson $B$ in
the SM. The hypercharge gauge field $B_\mu$ is decomposed into
a photon field $A_\mu$ and a $Z$-boson field $Z_\mu$ as
$B_\mu=c_W A_{\mu}-s_W Z_{\mu}$ with $c_W=\cos\theta_W$ and
$s_W=\sin\theta_W$ of the weak mixing angle $\theta_W$ after
the firmly-established EWSB. Furthermore, for simplicity
and without loss of generality, we omit the spin index $s$
of the Majorana particle $\chi_s$ in the following discussion,
as it applies universally across all spin cases.

\setcounter{equation}{0}

\section{DM relic abundance}
\label{sec:dm_relic_abundance}

In this section, we calculate the relic abundance of our anapole DM of
each spin from the thermal freeze-out mechanism.
The overabundant region beyond the observed relic abundance~\cite{Planck:2018vyg}
is simply considered to be excluded without introducing late time reduction
possibilities.
The corresponding areas are shown in the two-dimensional planes of mass 
of dark matter and the couplings $a_i, b_j$ normalized by the cutoff scale 
squared $\Lambda^2$.

\begin{figure}[ht!]
\vskip 0.5cm
\centering
\includegraphics[scale=1]{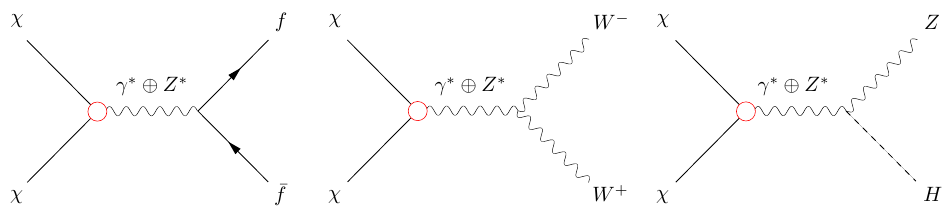}
\caption{\rm
    Feynman diagrams for the dominant annihilation processes
    of two identical Majorana particles into a pair of SM
    particles, $\chi\chi \rightarrow f\bar{f}$ (left),
    $W^-W^+$ (middle)
    and $ZH$ (right) where $f$ is  a SM quark $q=u,d,s,c,b,t$
    or lepton $\ell=e,\mu,\tau,\nu_e,\nu_\mu,\nu_\tau$.
    Here, $\chi$ denotes the self-annihilating Majorana DM particle. The red open circle in each diagram
    indicates the effective three-point anapole vertex. The notation
    $\gamma^*\oplus Z^*$ stands for the combined $s$-channel
    $\gamma$ and $Z$ exchanges.}
\label{fig:diagrams_annihilations}
\end{figure}

The self-conjugate hypercharge anapole DM particles can annihilate
into the SM particles via the $s$-channel photon $\gamma$
and $Z$ boson exchanges. If kinematically allowed,
the DM particles annihilate mainly via the processes
$\chi\chi \rightarrow f\bar{f}$,
$W^-W^+$ and/or $ZH$, where $f$ is a SM quark $q=u,d,s,c,b,t$
or lepton $\ell=e,\mu,\tau, \nu_e, \nu_\mu, \nu_\tau$, as depicted
in Fig.~\ref{fig:diagrams_annihilations}.
The relic abundance of $\chi$ can be determined through the freeze-out of the annihilation processes in the figure.
As noted previously in Ref.~\cite{Boudjema:1990st}, every
annihilation
cross section is completely factored into a simple product of
two independent parts, of which one corresponds to the DM
annihilation into a virtual gauge boson and the other to
the sequential decay of the virtual gauge
boson into a pair of SM particles.\footnote{The angular
distribution of each annihilation mode is uniquely determined
independently of the DM particle spin. This characteristic
spin-independent angular distribution was demonstrated
explicitly in
the scattering process
$e^-e^+\rightarrow\gamma^*\rightarrow \chi\chi$
via a photon exchange in Ref.$\,$\cite{Boudjema:1990st}.}
Explicitly, the total cross section of each annihilation
mode can be written in the following compact form:
\begin{align}
   \sigma_{1/2}
&= \frac{|a_{1/2}|^2}{4\Lambda^4}\, \beta_\chi\,s \,
   {\cal P}_{\scriptsize\mbox{SM}},
%
\label{eq:spin-1/2_annihilation_xsection}
\\
   \sigma_1
&=  \frac{|a_1|^2 \,\beta_\chi^2 + |b_1|^2\,}{9\Lambda^4}\,
    \left(\frac{s}{4m_\chi^2}\right)\, \beta_\chi\,s\,
   \,{\cal P}_{\scriptsize\mbox{SM}},
\label{eq:spin-1_annihilation_xsection}
\end{align}
for the spin-1/2 and spin-1 Majorana particles, respectively,
and
\begin{align}
   \sigma_{3/2}
&= \frac{|a_{3/2}|^2}{72\Lambda^4}\,
 \left(5-2\beta^2_\chi+5\beta^4_\chi\right)\,
 \left(\frac{s}{4m^2_\chi}\right)^2\,
\beta_\chi\,s \,
   {\cal P}_{\scriptsize\mbox{SM}},
\label{eq:spin-3/2_annihilation_xsection}
\\
   \sigma_2
&=  \frac{1}{300\Lambda^4}
\bigg[
|a_{2}|^2\beta_\chi^2
 \left(7-6\beta^2_\chi+15\beta^4_\chi\right)
+ |b_{2}|^2
\left(15-6\beta^2_\chi+7\beta^4_\chi\right)
\bigg]
    \left(\frac{s}{4m_\chi^2}\right)^3
    \, \beta_\chi\,s\,
   \,{\cal P}_{\scriptsize\mbox{SM}},
\label{eq:spin-2_annihilation_xsection}
\end{align}
for the spin-3/2 and spin-2 Majorana particles, respectively,
with the collision energy $\sqrt{s}$ and the
speed of $\chi$, $\beta_\chi=\sqrt{1-4m_\chi^2/s}$
in the center of mass (CM) frame
of two $\chi$ particles.
Here, the dimensionless SM pair-production term
${\cal P}_{\scriptsize\mbox{SM}} =
\sum_f {\cal P}_{f\bar{f}}+{\cal P}_{WW}+{\cal P}_{ZH}$ is
the sum of all the kinematically allowed production terms:
\begin{align}
   {\cal P}_{f\bar{f}}
&= \frac{e^2}{12\pi c_W^2} \beta_f\, \Pi_Z(s)
   \bigg[(3-\beta_f^2)\bar{V}_f^2
        +2\beta_f^2 A^2_f\bigg]\,\theta(\sqrt{s}-2m_f)\,,
\label{eq:2-body_decay_ff}
\\
    {\cal P}_{WW}
&= \frac{e^2}{96\pi c_W^2}
   (1+\Gamma_Z^2/m_Z^2) \, \beta_W^{3}\, \Pi_Z(s)\,
   (1+20 m_W^2/s+12m_W^4/s^2)\,
         \theta(\sqrt{s}-2 m_W)\,,
\label{eq:2-body_decay_ww}
\\
    {\cal P}_{ZH}
&= \frac{e^2 }{96\pi c_W^2}\, \bar{\beta}_{ZH}\,\Pi_Z(s)\,
   (1+8m_Z^2/s)\,
   \theta(\sqrt{s}-m_Z-m_H)\,,
\label{eq:2-body_decay_zh}
\end{align}
with $\beta_{f,W}=\sqrt{1-4m_{f,W}^2/s}$ and
$\bar{\beta}_{ZH}=\sqrt{[1-(m_Z+m_H)^2/s][1-(m_Z-m_H)^2/s]}$.
Here, for the sake of notation, the normalized propagator factor
$\Pi_Z(s)$ and an effective SM vector coupling squared
$\bar{V}^2_f$ are introduced as
\begin{align}
   \Pi_Z(s)
&= \frac{s^2}{(s-m_Z^2)^2+m_Z^2\Gamma_Z^2}\,,
\label{eq:normalized_propagator_factor}
   \\
   \bar{V}^2_f(s)
&= V_f^2-2c_W^2Q_f V_f
         \bigg(1-\frac{m_Z^2}{s}\bigg)
           +c_W^4\frac{Q_f^2}{\Pi_Z(s)}\,,
\label{eq:effective_vector_coupling_square}
\end{align}
in terms of the SM vector and axial-vector couplings,
$V_f=I^3_f/2-Q_f s_W^2$
and $A_f=-I^3_f/2$, of the $Z$ boson to a SM fermion pair $f\bar{f}$
with the isospin component $I^3_f$ and electric charge
$Q_f$ of the fermion $f$. We emphasize
again that the SM production terms are independent of the DM
particle spin
and its couplings to the photon and $Z$ boson.
Consequently, the information on the characteristics of the
anapole DM particle is encoded exclusively in the DM
annihilation into a virtual gauge boson.

\begin{figure}[ht!]
\centering
\includegraphics[scale=0.5]{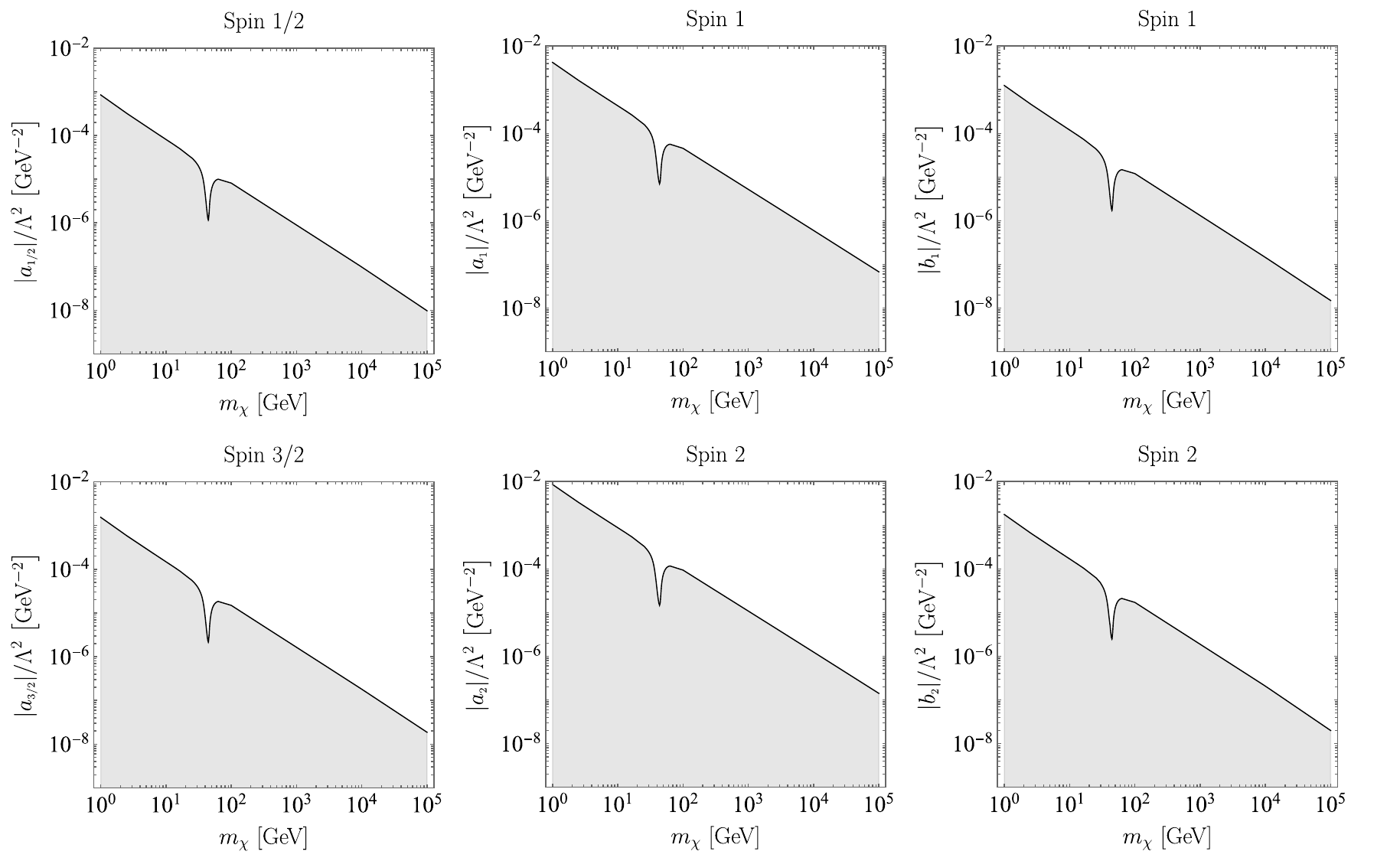}
\caption{\rm
    Exclusion limits on the effective anapole
    couplings versus the DM mass from the observed DM relic
    abundance.
    The top (bottom) left panel shows the constraint
    on the normalized coupling $|a_{1/2}|/\Lambda^{2}$
    ($|a_{3/2}|/\Lambda^{2}$) in the
    spin-1/2 ($3/2$) case. The top (bottom) middle panel
    shows the constraint on the normalized coupling
    $|a_1|/\Lambda^{2}$ ($|a_{2}|/\Lambda^{2}$) and
    the top (bottom) right panel shows the constraint
    on the normalized couplings $|b_1|/\Lambda^{2}$
    ($|b_{2}|/\Lambda^{2}$) in the spin-1 ($2$) case.
    In each plot, the grey-shaded region is excluded 
    by making DM overabundant and bounded by 
    the relic density line (black solid).
    }
\label{fig:relic_plots}
\end{figure}

By calculating the DM relic abundance
using the methodology described in detail
in Appendix~\ref{appendix:relic_calculation_strategy},
we can derive exclusion limits on the effective anapole
couplings versus the DM mass. These limits are based on
the observed relic abundance of
$\Omega_\chi h^2\approx 0.12$, as illustrated in
Fig.\ref{fig:relic_plots}. Each plot features a grey
shaded region that represents an overabundance of DM and
is bounded by the relic density line (black solid) obtained
from the freeze-out mechanism.
The top (bottom) left panel is for the normalized coupling
$|a_{1/2}|/\Lambda^{2}$ ($|a_{3/2}|/\Lambda^{2}$) versus
the DM mass $m_\chi$ in
the spin-1/2 (3/2) case, the top (bottom) middle  panel is for
the normalized coupling
$|a_1|/\Lambda^{2}$ ($|a_2|/\Lambda^{2}$) versus the DM mass
$m_\chi$ in the spin-1 (2) case, and
the top (bottom) right  panel is for
the normalized coupling
$|b_1|/\Lambda^{2}$ ($|b_2|/\Lambda^{2}$) versus the DM mass
$m_\chi$ in the spin-1 (2) case.
The single-power dependence of the fermionic
annihilation cross sections on the $\chi$ speed $\beta_\chi$ in
Eq.$\,$\eqref{eq:spin-1/2_annihilation_xsection}
implies that each DM annihilation is a $p$-wave
dominant process in the fermionic cases.
On the other hand, the bosonic DM annihilation cross sections
include the $d$-wave dominant terms proportional to the coupling
$a_1$ or $a_2$ in addition to the $p$-wave dominant ones with
the coupling $b_1$ or $b_2$. Hence, the cross sections can be
further suppressed by $\beta_\chi^4$ once a UV model expects
$b_1$ or $b_2$ is negligible, as can be clearly seen in the
right panels of Fig.~\ref{fig:relic_plots}.
Note that the $p$-wave dominant terms $|b_{1,2}|/\Lambda^{2}$ are more strongly
constrained than the $|a_{1/2,3/2}|/\Lambda^{2}$ from the observed relic
abundance due to the reduced spin-averaged and
polarization-weighted factors as
shown in Eqs. (\ref{eq:spin-1/2_annihilation_xsection})
and (\ref{eq:spin-1_annihilation_xsection}).
Typically, due to the smaller spin averaged factors in
the annihilation cross sections, higher-spin DM particles
face more stringent constraints compared to the lower-spin
cases with the same order of suppression factor $\beta_\chi$.

\setcounter{equation}{0}

\section{LHC searches}
\label{sec:collider_searches}

In this section, we derive the exclusion limits on the couplings
for the hypercharge anapole DM particle from its searches at
the LHC experiment with the projected sensitivities
on the couplings from the upcoming HL-LHC
experiment~\cite{ATLAS:2021kxv,CMS:2021far,
Chakraborty:2018kqn,Frattari:2020tiy}.
Although there are possibly various production channels at the LHC, we consider the most dominant production
processes for the hypercharge anapole DM particle in
 our analytic analysis.

It was shown in a previous work~\cite{Florez:2019tqr} that
the EM anapole DM particles in the U(1)$_{\rm EM}$ gauge-invariant
framework can be produced dominantly through the di-jet
processes via vector-boson fusion, especially when investigating
them with strong experimental cuts. On the contrary,
the di-jet processes cannot be dominant anymore in the hypercharge
anapole DM case because
not only the $\gamma$ exchange diagram but also $Z$-boson exchange
diagram contribute to the process, leading to a quite significant
cancellation in the high-energy regime so that the unitarity problem
is diminished extremely efficiently~\cite{Arina:2020mxo}.
As a result, the strongest LHC constraints on
the hypercharge anapole DM couplings are
expected to come from the so-called
mono-jet processes $pp\to j+X$ with $X$ standing for
the collection of invisible particles including two DM
particles, $\chi\chi$.\s

\begin{figure}[ht!]
\centering
\includegraphics[scale=1.1]{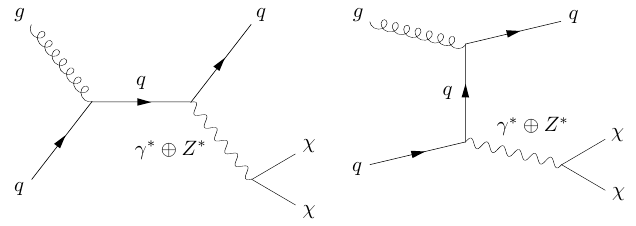}
\caption{\rm
     Two parton-level Feynman diagrams contributing to
     the process
     $gq\to q\chi\chi$, of which the left one is for a $s$-channel
     quark exchange and the right one is for a $t$-channel
     quark exchange.
     If the quark mass is ignored, the $t$-channel diagram has a
     forward singularity to be regularized. The notation
    $\gamma^*\oplus Z^*$ stands for the combined $s$-channel
    $\gamma$ and $Z$ exchanges. }
\label{fig:parton_level_mono-jet_processes}
\end{figure}

Two parton-level processes, $gq\to q\chi\chi$ and
$q\bar{q}\to g \chi\chi$, contribute dominantly to
the mono-jet process $pp\to j+X$ at the LHC.
Quantitatively, the former $gq$ cross section
is much larger than the latter $q\bar{q}$ cross section.
In this light, we consider the process $gq\to q\chi\chi$
in the present work as the most crucial mode for investigating
the constraints from the LHC mono-jet events on the couplings
versus the DM mass $m_\chi$ in the spin-1/2, 1, 3/2,
and 2 cases.\footnote{The parton-level process
$g\bar{q}\to \bar{q} \chi\chi$ contributes to the same
single-jet process at the LHC, although this mode is insignificant
because the anti-quark
contribution is much smaller than the quark contribution to
the parton distribution functions of the proton.}
There are two Feynman diagrams contributing to the sequential
process $gq\to q \,\gamma^*/Z^*\to q \chi\chi$,
of which one is a $s$-channel quark-exchange mode and
the other a $t$-channel quark-exchange
mode as depicted in Fig.~\ref{fig:parton_level_mono-jet_processes}.
As demonstrated in all the
DM annihilation processes in Sect.$\,$\ref{sec:dm_relic_abundance},
the effective anapole three-point $\gamma\chi\chi$ and $Z\chi\chi$
vertices allow each parton-level cross section to be completely
factored into the SM 2-to-2 scattering process
$gq\to q\gamma^*/Z^*$ and
the DM pair production processes through the 2-body decay
$\gamma^*/Z^*\to \chi\chi$. After performing the 2-body
phase space integration over the invisible final two-body
$\chi\chi$ system, we can obtain the parton-level cross-section for
the process $gq \rightarrow q \chi\chi$ in a compact integral
form as
\begin{align}
   \sigma_{1/2}(\hat{s};\,p_{_T})
&= \int^{\hat{s}}_{4m_\chi^2} \frac{dQ^2}{2\pi}
   \mathcal{O}(\hat{s},Q^2;\,p_{_T})\,
   \frac{|a_{1/2}|^2}{\Lambda^4},
\label{eq:spin-1/2_parton_level_process}
\\
   \sigma_{1}(\hat{s};\, p_{_T})
&= \int^{\hat{s}}_{4m_\chi^2}
    \frac{dQ^2}{2\pi} \mathcal{O}(\hat{s},Q^2;\, p_{_T})\,
    \left(\frac{Q^2}{4m_\chi^2}\right)
   \frac{|a_1|^2\,\beta_\chi^2 + |b_1|^2}{\Lambda^4},
\label{eq:spin-1_parton_level_process}
\\
   \sigma_{3/2}(\hat{s};\,p_{_T})
&= \int^{\hat{s}}_{4m_\chi^2} \frac{dQ^2}{2\pi}
   \mathcal{O}(\hat{s},Q^2;\,p_{_T})\,
   \left(\frac{Q^2}{4m^2_\chi}\right)^2
   \frac{2|a_{3/2}|^2}{9\Lambda^4}
\left(5-2\beta^2_\chi+5\beta^4_\chi\right),
\label{eq:spin-3/2_parton_level_process}
\\
   \sigma_{2}(\hat{s};\, p_{_T})
&= \int^{\hat{s}}_{4m_\chi^2}
    \frac{dQ^2}{2\pi} \mathcal{O}(\hat{s},Q^2;\, p_{_T})\,
    \left(\frac{Q^2}{4m_\chi^2}\right)^3\,
   \nonumber\\
 &\qquad\times
 \frac{1}{12\Lambda^4}
\bigg[
|a_{2}|^2 \beta_\chi^2
\left(7-6\beta^2_\chi+15\beta^4_\chi\right)
+ |b_{2}|^2
\left(15-6\beta^2_\chi+7\beta^4_\chi\right)
\bigg],
\label{eq:spin-2_parton_level_process}
\end{align}
for the spin-1/2, 1, 3/2, and 2 cases, respectively,
with the $\chi\chi$ invariant mass $\sqrt{Q^2}$ corresponding to the
virtual $\gamma^*$ or $Z^*$ invariant mass, the speed factor
$\beta_\chi=\sqrt{1-4m_\chi^2/Q^2}$  as well as the $gq$ collision
CM energy $\sqrt{\hat{s}}$ and the transverse-momentum cut $p_{_T}$
of the produced quark, invariant under any Lorentz boost
along the proton beam direction.  The parton-level 2-to-2
scattering processes
$gq\rightarrow q\gamma^*/Z^*$ are encoded fully in the $\hat{s}$ and
$Q^2$ dependent cross section $\mathcal{O}$. Explicitly the
parton-level effective cross section ${\cal O}$ with the
transverse-momentum cut $p_{_T}$ is given by
\begin{align}
    \mathcal{O}(\hat{s},Q^2;\, p_{_T})
=  \frac{e^2g_S^2}{192\pi^2 c_W^2\hat{s}}
    (|\bar{V}_q|^2+A_q^2\,)\,Q^2\Pi_Z(Q^2)
    \bigg(1-\frac{4m_\chi^2}{Q^2}\bigg)^{3/2}\,
    {\cal F}(\hat{s}, Q^2;\,p_{_T}),
\label{eq:gq_to_qV}
\end{align}
with the strong-interaction coupling $g_s$, the normalized
propagator
$\Pi_Z(Q^2)$ and the modified vector coupling $\bar{V}_q$ of the
quark $q$  defined in Eqs.~\eqref{eq:normalized_propagator_factor}
and~\eqref{eq:effective_vector_coupling_square}, where
the parton-level
$q$ transverse momentum, which is invariant under the Lorentz boost
along the beam direction, is
$\hat{p}_{_T}= \frac{\sqrt{\hat{s}}}{2}(1-Q^2/\hat{s})\,\sin\theta$
with the polar-angle $\theta$ between the momenta of
the initial gluon $g$ and the final quark $q$.
The function ${\cal F}(\hat{s}, Q^2;\,p_{_T})$ with the
transverse-momentum
cut $p_{_T}$ in Eq.~\eqref{eq:gq_to_qV} is given explicitly by
\begin{eqnarray}
   {\cal F}(\hat{s}, Q^2;\,p_{_T})
&=&  \bigg(\frac{1}{3\sqrt{{\hat{s}}}}+\frac{Q^2}{\hat{s}^{3/2}}\bigg)
   \sqrt{{p^2_{_{T \rm max}}}-p^2_{_T}}
  +\bigg(\frac{4 p^2_{_{T \rm max}}}{3\hat{s}}
        +\frac{Q^4}{3{\hat{s}^2}}\bigg)\,
         \ln \left(\frac{p_{_{T \rm max}}
                        +\sqrt{p_{_{T \rm max}}^2-p^2_{_T}}}
                        {p_{_T}}\right)
  \nonumber\\
&\sim & \frac{1}{6} \left(1+2 \frac{Q^2}{\hat{s}}
                            - 3\frac{Q^2}{\hat{s}}\right)
         +\frac{1}{3} \left(1-2\frac{Q^2}{\hat{s}}
                             +2\frac{Q^4}{\hat{s}^2}\right)
         \ln \left(\frac{\hat{s}-Q^2}{\sqrt{\hat{s}}p_{_T}}
             \right) \ \ \mbox{for}\ \ p_{_T}\to 0,
\label{eq:p_T-dependent_distribution}
\end{eqnarray}
with the maximal transverse momentum
$p_{_{T \rm max}}= \frac{\sqrt{\hat{s}}}{2}(1-Q^2/\hat{s})$.
The transverse-momentum cut $p_{_T}$
is introduced to regularize the forward singularity
caused by neglecting the quark mass; this also allows us
to ignore particles escaping detection
along the proton-beam pipe directions.
Compared to the spin-1/2 case, the parton-level production
cross section in the spin-1 case has an additional
kinematic enhancement factor
$Q^2/4m_\chi^2$ originating from the longitudinal mode of
one of the two spin-1 DM particles
as in Eq.~\eqref{eq:spin-1_parton_level_process}.
We note in passing that the power of the enhancement factor gets
bigger for higher spin cases
due to the larger number of longitudinal modes so that
the mono-jet searches at the LHC impose much stronger constraints
on the couplings gradually.
Certainly, the parton-level cross sections
should be folded with proper quark and gluon parton distribution
functions for evaluating the mono-jet cross section at the LHC.

In this paper, we adopt a simple statistical analysis
for deriving the LHC and HL-LHC limits on the hypercharge
anapole couplings versus the DM mass, focusing on the effective
characterization of the
hypercharge anapole DM particle according to the spin. Let us
calculate
the simplest version of the signal significance $z$ defined by
\begin{align}
z=\frac{s}{\sqrt{s+b}},
\end{align}
with the numbers $s$ and $b$ of the signal and background
events.\footnote{The most significant parton-level background
process is
$gq\to q\nu\bar{\nu}$ with two invisible neutrinos in the final
state produced via a $Z$-boson exchange.}
We set the critical significance value $z=2$ to determine the
95\% confidence
level (CL) exclusion limits on the couplings normalized to the
cutoff scale squared $\Lambda^2$. The events are selected according
to the selection criteria $p_T^{jet}>250$ GeV and $|\eta|<2.5$,
applied to the most recent mono-jet searches with an integrated
luminosity of 139 fb$^{-1}$ at the LHC energy of 13 TeV by
the ATLAS collaboration in Ref.~\cite{ATLAS:2021kxv}.

\begin{figure}[ht!]
\centering
\includegraphics[scale=0.48]{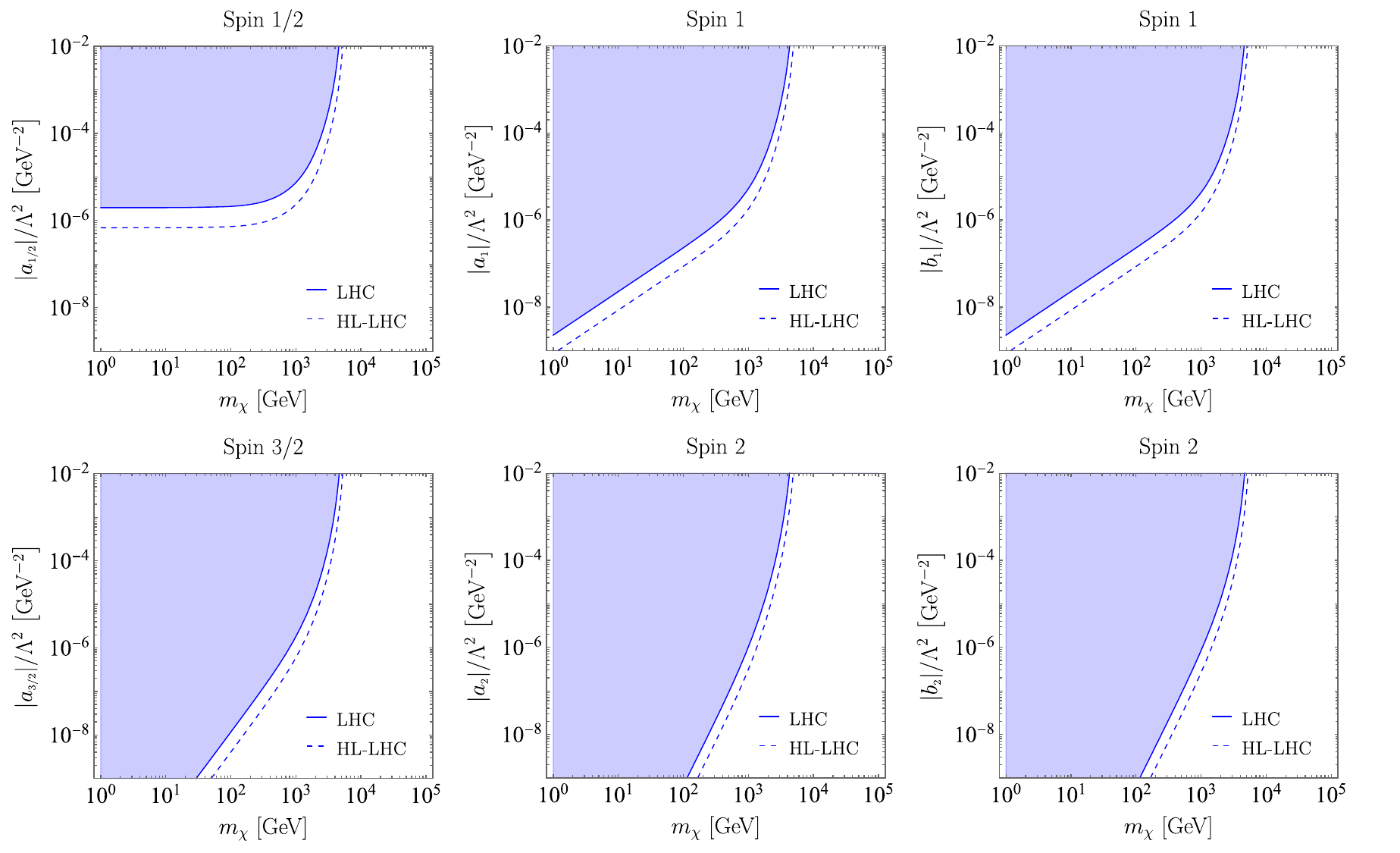}
\caption{\rm
    Exclusion limits from the mono-jet events at the LHC
    (solid line) of 13 TeV energy and 139 fb${}^{-1}$ luminosity
    and projected sensitivities from those at the HL-LHC (dashed
    line) of 14 TeV energy and 3 ab${}^{-1}$ luminosity,
    respectively, derived mainly from the parton-level process
    $gq\to q\chi\chi$,  on the effective couplings
    and the DM mass
    $m_\chi$.
    The top (bottom) left panel shows the limits on the normalized
    coupling $a_{1/2}/\Lambda^{2}$ ($a_{3/2}/\Lambda^{2}$) in the
    spin-1/2 (3/2) case. The top (bottom) middle panel shows the
    limits on the normalized coupling  $a_1/\Lambda$ ($a_2/\Lambda^2$)
    and the top (bottom) right panel shows the
    limits on the normalized coupling  $b_1/\Lambda^2$ ($b_2/\Lambda^2$) in the spin-1 (2) case.
    }
\label{fig:collider_plots}
\end{figure}

Figure~\ref{fig:collider_plots} show the current exclusion limits from the LHC
(shaded region bounded by the solid lines) and the expected sensitivities
at the HL-LHC with the full running of the 3 ab$^{-1}$ integrated luminosity
(dashed lines).
Comparing the panels from the top left one, it is clearly seen that
the couplings in the higher-spin cases are constrained more strongly
in the whole kinematically-available region due to the higher power of
the enhancement factor $Q^2/4m^2_\chi$, especially in the low mass region
as indicated in Eqs.~(\ref{eq:spin-1_parton_level_process}),
(\ref{eq:spin-3/2_parton_level_process}) and (\ref{eq:spin-2_parton_level_process}).

Close to the kinematical endpoint of
the mass $m_\chi \sim 6\, {\rm TeV}$,
there is a slight reduction in the
$a_1/\Lambda^2$ ($a_2/\Lambda^2$) constraint compared
to the $b_1/\Lambda^{2}$ ($b_2/\Lambda^2$) constraint, due to the presence of
the higher power of the kinematical suppression factor $\beta^2_\chi=1-4m^2_\chi/Q^2$
for $Q^2\sim 4 m_\chi^2$ in the spin-1 (2) case.

The dashed line in each panel shows the future sensitivity of
the planned HL-LHC experiment with a slightly larger collision
energy of 14 TeV and an integrated luminosity of 3 ab$^{-1}$,
roughly 10 times larger than the present LHC
luminosity~\cite{Shiltsev:2019rfl,Assmann:2023dwx}.
We require slightly stronger selection criteria of the
mono-jet events for the HL-LHC: $p_T^{jet}>300$ GeV and $|\eta|<2.5$.
As for the number of background events at $\sqrt{s} = 14$ TeV,
we consider the difference in the cross sections of the dominant
background process $pp\rightarrow Zj\to \nu\bar{\nu}\, j$,
enhanced by $11.57/9.88\approx 1.17$ compared to the
$\sqrt{s} = 13$ TeV case~\cite{Chakraborty:2018kqn}.

The projected sensitivities become stronger as $m_\chi$ decreases.
Due to the $Q^2$ dependence of the $gq \to q \chi \chi$ production cross section,
we expect that the 100 TeV future circular collider experiment under research
and development (R\&D)~\cite{FCC:2018vvp,Benedikt:2020ejr} enables
us to cover a much larger region of the couplings versus the anapole DM mass.

Before closing this section, we emphasize
once more that the power of the invariant mass square
in the mono-jet cross section increases in
proportion to the spin value of the anapole DM particle.
The enhancement arises from the increased number of longitudinal
modes of the higher-spin anapole DM particle.
It strongly indicates that the constraints on the couplings
versus the DM mass from the LHC and HL-LHC mono-jet
searches become much stronger as the spin value of
the anapole DM particle increases.

\setcounter{equation}{0}

\section{DM direct detection}
\label{sec:direct_searches}

In this section, we describe how the exclusion limits on the
hypercharge anapole DM couplings versus the DM mass are
extracted from the recent DM direct detection experiment XENONnT
with the 1.1 ton-year exposure~\cite{XENON:2023cxc}
which is at present the most powerful DM direct detection
experiment. Along with the exclusion limits,
we consider the projected sensitivities
of the future XENONnT with the 20 ton-year exposure.
Non-relativistic dark matter is assumed to move with a typical velocity of order $v\simeq 10^{-3}c$ in the galactic
halo.\footnote{Note that the scenarios with fast-moving light DM,
so-called {\it Boosted Dark Matter}, are proposed in Refs.~\cite{Agashe:2014yua,Kim:2016zjx,Giudice:2017zke,Alhazmi:2020fju} but they are not the majority of the cosmological DM with the observed relic abundance.}
Thus, the recoil energy of a DM particle against a heavy
target nucleus is expected to be in the keV energy scale, much
smaller than the typical DM mass
$\gtrsim \mathcal O ({\rm GeV})$ in consideration here
as well as the $Z$-boson mass
$m_Z= 91.2\, {\rm GeV}$. Hence, the $Z$-boson exchange contribution
to the elastic DM scattering off the target nucleus can be safely
ignored because the momentum transfer is significantly
smaller than the $Z$ boson mass $m_Z$
and the DM scattering off the nucleus is dominated by the photon exchange~\cite{Arina:2020mxo}.

Taking the small recoil energy limit and ignoring the $Z$-boson
exchange contribution safely we can cast the recoil-energy
dependent differential cross section into the following
factorized form:
\begin{align}
   \frac{d\sigma_{1/2}}{dE_R}
&= \frac{c_W^2e^2}{4\pi^2}\frac{|a_{1/2}|^2}{\Lambda^4}\,
   \mathcal{A}(E_R)\,,
\label{eq:spin-1/2_recoil_energy_distribution}
\\
   \frac{d\sigma_{1}}{dE_R}
&= \frac{c_W^2e^2}{6\pi^2}\frac{1}{\Lambda^4}
   \bigg[|a_{1}|^2\bigg(1+\frac{m_T E_R}{2m_\chi^2}\bigg)
        +|b_{1}|^2\frac{m_T E_R}{2m_\chi^2}
   \bigg]\mathcal{A}(E_R)\,,
\label{eq:spin-1_recoil_energy_distribution}
\\
   \frac{d\sigma_{3/2}}{dE_R}
&= \frac{5c_W^2e^2}{36\pi^2}\frac{|a_{3/2}|^2}{\Lambda^4}\,
   \mathcal{A}(E_R)\,,
\label{eq:spin-3/2_recoil_energy_distribution}
\\
   \frac{d\sigma_{2}}{dE_R}
&= \frac{c_W^2e^2}{120\pi^2}\frac{1}{\Lambda^4}
   \bigg[15|a_{2}|^2\bigg(1+\frac{13m_T E_R}{10m_\chi^2}\bigg)
        +7|b_{2}|^2\frac{m_T E_R}{2m_\chi^2}
   \bigg]\mathcal{A}(E_R)\,,
\label{eq:spin-2_recoil_energy_distribution}
\end{align}
for the spin-1/2, 1, 3/2, and 2 cases, respectively, with the
target nucleus mass $m_T$ and the recoil energy $E_R$.
Note that the momentum transfer
$q = \sqrt{2m_TE_R}$ is much smaller than our dark matter mass in consideration.
One noteworthy feature is that the recoil-energy dependent function $\mathcal{A}
(E_R)$ is factored out independently of the spin of the anapole
DM particle:
\begin{align}
  \mathcal{A}(E_R)
= Z^2_T\bigg[2m_T-\bigg(1+\frac{m_T}{m_\chi}\bigg)^2
  \frac{E_R}{v_\chi^2}\bigg] F^2_C(q^2)
 +\frac{2m_T^2}{3m_p^2}\bigg(\frac{\bar{\mu}_T}{\mu_p}\bigg)^2
  \frac{E_R}{v_\chi^2}
  F^2_M(q^2)\,,
\label{eq:a_er_function}
\end{align}
with the atomic number
$Z_T$ of the target nucleus and
the DM particle speed $v_{\chi}$ relative to the nucleus.
The charge form factor $F_C$ is given by
\begin{align}
   F_C(q^2)
= \bigg(\frac{3\, j_1(qr_C)}{qr_C}\bigg)e^{-q^2s^2/2}\,,
\label{eq:function_f_c}
\end{align}
in terms of the first-kind spherical Bessel function $j_1$ of
order 1 where $r_C=(c^2+7\pi^2a^2/3-5s^2)^{1/2}$ with
$c\simeq (1.23 A^{1/3}- 0.60)\,{\rm fm}$,
$a\simeq 0.52\,{\rm fm}$ and $s\simeq 0.9\,{\rm fm}$, and
the atomic mass
$A$ of the target nucleus, while the magnetic
dipole moment form factor
$F_M$~\cite{Helm:1956zz,Lewin:1995rx} is given by
\begin{align}
   F_M(q^2)
= \left\{\begin{array}{ll}
         \displaystyle
              \frac{\sin(qr_M)}{qr_M}&
              \mbox{for }qr_M<2.55,\; qr_M>4.5\,,
              \\[15pt]
              0.2168&\mbox{for } 2.55<qr_M<4.5\,,
\end{array}
\right.
\label{eq:function_f_m}
\end{align}
with the radius $r_M=1.0 A^{1/3}\,\, {\rm fm}$.
The recoil-energy dependent function ${\cal A}$
in Eq.$\,$(\ref{eq:a_er_function}) involves the nuclear magneton
$\mu_p=e/2m_p$ with the proton mass $m_p$ and the weighted dipole
moment $\bar{\mu}_T$ for the target nuclei~\cite{Chang:2010en}:
\begin{align}
\bar{\mu}_T=\bigg(\sum_i f_i\mu_i^2 \frac{s_i+1}{s_i}\bigg)^{1/2}\,,
\end{align}
where $f_i$, $\mu_i$, and $s_i$ are the abundance
fraction, magnetic moment, and spin of the isotope $i$.

The recoil-energy dependent distribution of the DM direct detection
process is given by integrating the differential cross section
of each DM spin as in Eqs.~\eqref{eq:spin-1/2_recoil_energy_distribution} -
\eqref{eq:spin-2_recoil_energy_distribution} over the DM velocity with the
distribution $f_{\tiny\mbox{LAB}}(\vec v\,)$ in
the laboratory frame as
\begin{align}
  \frac{dR}{dE_R}
= \frac{1}{m_T}\frac{\rho_{\scriptsize\mbox{loc}}}{m_\chi}
   \int d^3 \vec v \;|\vec{v}\,|\, f_{\tiny\mbox{LAB}}(\vec v\,)
   \frac{d\sigma}{dE_R}\,,
\label{eq:recoil_energy_dm_detection_distribution}
\end{align}
where we use the local DM density
$\rho_{\scriptsize\mbox{loc}}=0.3$ GeV cm${}^{-3}$.
In the present work, we adopt a simple Maxwell-Boltzmann
distribution in the galactic frame truncated at the escape
speed $v_{\scriptsize\mbox{esc}}$
of our Galaxy:
\begin{align}
f_{\tiny\mbox{LAB}}(\vec v\,)=f(\vec v+\vec v_{\scriptsize\mbox{E}})\,,
\end{align}
with the velocity $\vec v_{\scriptsize\mbox{E}}$ of the Earth
in the galactic frame and
\begin{align}
  f(\vec v\,)
= \left\{
  \begin{array}{ll}
  \displaystyle
  \frac{1}{\mathcal{N}} e^{-v^2/v_0^2} & \mbox{ for }|\vec v\,|\leq
   v_{\scriptsize\mbox{esc}}\,,
  \\[10pt]
   0 & \mbox{ for }|\vec v\,|>v_{\scriptsize\mbox{esc}}\,,
\end{array}
\right.
\label{eq:function_f_v}
\end{align}
where the normalization constant ${\cal N}$ is
\begin{align*}
{\cal N}=(\pi v^2_0)^{3/2}\bigg[\mbox{erf}
\bigg(\frac{v_{\scriptsize\mbox{esc}}}{v_0}\bigg)
-\frac{2}{\sqrt{\pi}}
\frac{v_{\scriptsize\mbox{esc}}}{v_0}
e^{-v^2_{\scriptsize\mbox{esc}}/v_0^2}\bigg]\,,
\end{align*}
with the error function
${\rm  erf}(z)=2\int^z_0 e^{-t^2} dt/\sqrt{\pi}$.
The values of the three different speeds are set numerically
to the escape speed
$v_{\scriptsize\mbox{esc}}=544$ km s${}^{-1}$,
the speed of the Sun relative to the DM reference frame
$v_{0}=220$ km s${}^{-1}$ and the Earth speed
$v_{E}=232$ km s${}^{-1}$ in the galactic frame.

\begin{figure}[ht!]
\centering
\includegraphics[scale=0.5]{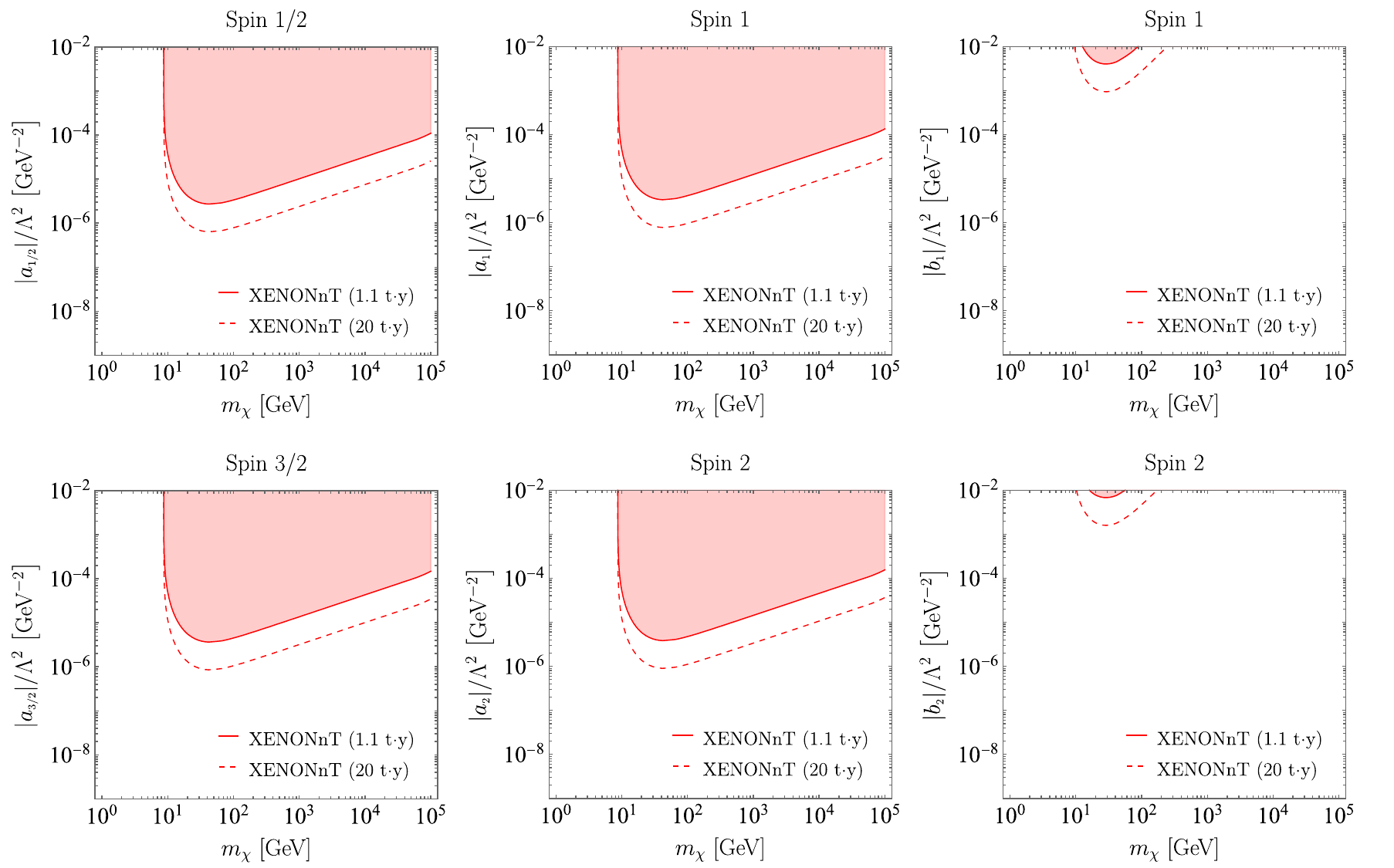}
\caption{\rm Exclusion limits from the
    DM direct detection experiment
    XENONnT on the effective couplings versus
    the DM mass $m_\chi$.
    The solid line is the current limit
    with the 1.1  ton-year exposure and the dashed line is the expected
    sensitivity with the 20 ton-year exposure.
    The top (bottom) left panel
    shows the limit on the normalized coupling
    $|a_{1/2}|/\Lambda^{2}$ ($|a_{3/2}|/\Lambda^{2}$)
    versus the DM mass in the spin-1/2 (3/2) case.
    The top (bottom) middle panel
    shows the limit on the normalized coupling
    $|a_1|/\Lambda^2$ ($|a_2|/\Lambda^{2}$) versus the DM mass
    in the spin-1 (2) case while setting the other coupling to zero.
    The top (bottom) right panel
    show the limit on the normalized coupling
    $|b_1|/\Lambda^2$ ($|b_2|/\Lambda^{2}$) versus the DM mass
    in the spin-1 (2) case while setting the other coupling to zero.
    The right panels clearly show that the
    the normalized couplings, $|b_1|/\Lambda^{2}$ and $|b_2|/\Lambda^{2}$,
    for the spin-1 and spin-2 DM cases get much weaker constraints than
    the other cases.
    }
\label{fig:direct_plots}
\end{figure}

By integrating out the recoil-energy distribution
$dR/dE_R$ in Eq.$\,$(\ref{eq:recoil_energy_dm_detection_distribution}) and
taking into account the detection efficiencies of the
XENONnT experiment~\cite{XENON:2023cxc}, we can
evaluate the number of recoil DM detection events.
For the  details of calculating the expected signal events we refer to Appendix B
of Ref.~\cite{Hisano:2020qkq}.
The top (bottom) left panel of Fig.$\,$\ref{fig:direct_plots}
shows the 90\% CL
constraint on the normalized coupling
$|a_{1/2}|/\Lambda^{2}$ ($|a_{3/2}|/\Lambda^{2}$) versus
the DM mass
$m_\chi$ in the spin-1/2 (3/2) case. The top (bottom) middle panel
shows the constraint on the normalized coupling $|a_1|/\Lambda^{2}$
($|a_2|/\Lambda^{2}$) versus $m_\chi$, and the top (bottom) right panel
shows the constraint on the normalized coupling $|b_1|/\Lambda^{2}$
($|b_2|/\Lambda^{2}$) versus $m_\chi$ in the spin-1 (2) case.
In each plot, the excluded region of the corresponding coupling versus
the DM mass is shown as the red-shaded area bounded by
the exclusion limit with a red solid line.
Note that each of the second terms proportional to the
recoil energy $E_R$ on the couplings, $|a_1|/\Lambda^{2}$ and
$|a_2|/\Lambda^{2}$ in
Eqs.~\eqref{eq:spin-1_recoil_energy_distribution}
and~\eqref{eq:spin-2_recoil_energy_distribution}
is much smaller than unity, i.e.,
$m_TE_R/(2m_\chi^2)\ll 1$,
and hence the direct detection bounds on the $|a_i|/\Lambda^2$
couplings with $i=1/2,1,3/2,2$ are dominantly
determined by the spin-averaged factors.
On the other hand, the direct detection cross sections are
suppressed by $m_T E_R/2 m_\chi^2$ once they are dominated by
the $|b_i|/\Lambda^2$ term with $i=1,2$, which results in the
negligible sensitivities as shown in the right panels.
The tiny difference between the limits on the couplings,
$|b_1|/\Lambda^{2}$ and $|b_2|/\Lambda^{2}$, arise from
the different spin-averaged and polarization-weighted
factors, 1/6 and 7/120, in the detection rates, respectively.
For the expected sensitivities of the upcoming XENONnT with the
20 ton-year exposure shown with
the dashed lines, we relied simply on scaled statistics without accounting
for potential future
improvements in background rejection and the control of systematic uncertainties.

As shown previously in Fig.$\,\ref{fig:collider_plots}$,
the LHC and HL-LHC mono-jet constraints
on the couplings of a lower-spin particle
are much weaker than those on the couplings of
a higher-spin case, especially for the DM mass
$m_\chi\lesssim 1$ TeV.
Hence, the LHC/HL-LHC and DM direct detection experiments
can play quite complementary roles in imposing the exclusion limits on
the couplings versus the DM mass.

\section{Combined constraints and future sensitivities}
\label{sec:combined_constraints}

In this section, we show the results combining all the aforementioned 
experimental constraints and future sensitivities on the effective
couplings versus the DM mass coming from the Planck determination of the
DM relic abundance, the LHC and HL-LHC mono-jet searches, 
and the present DM direct detection experiment XENONnT and
its future data of 20 ton$\cdot$yr exposure, which have been
evaluated systematically in the previous three sections.
On top of those experimental bounds and sensitivities, we
include an additional theoretical constraint from the naive
perturbativity bound (NPB) for guaranteeing the validity of the
EFT formalism, which needs to be taken with a grain
of salt.\footnote{As the anapole terms are described
by an effective Lagrangian with higher-dimensional terms,
the so-called tree-level unitarity is violated in the high-energy
regime as well in various processes such as
$\chi\chi\to W^-W^+$. However, quantitatively the combined
constraint from the tree-level unitarity condition turns out to
be much weaker than that from the naive perturbativity
condition.} The energy-dependent NPBs on the couplings
simply read
\begin{align}
     \frac{|a_{1/2}|}{\Lambda^2}s
\leq 4\pi, \quad
     \frac{\sqrt{|a_1|^2 + |b_1|^2}}{\Lambda^2} s
\leq 4\pi\,, \quad
\frac{|a_{3/2}|}{\Lambda^2}s
\leq 4\pi, \quad\mbox{and}\quad
     \frac{\sqrt{|a_2|^2 + |b_2|^2}}{\Lambda^2} s
\leq 4\pi\,,
\label{eq:naive_perturbativity_condition}
\end{align}
for the spin-1/2, 1, 3/2, and 2 cases, respectively.
As the CM energy $\sqrt{s}\geq 2m_\chi$,
the NPB condition \eqref{eq:naive_perturbativity_condition}
applied to the asymptotically high-energy limit leads to
the following inequality relations:
\begin{align}
     \frac{|a_{1/2}|}{\Lambda^2}
\leq \frac{\pi}{m^2_\chi},
     \quad
     \frac{\sqrt{|a_1|^2 + |b_1|^2}}{\Lambda^2}
\leq \frac{\pi}{m^2_\chi}\,,\quad
 \frac{|a_{3/2}|}{\Lambda^2}
\leq \frac{\pi}{m^2_\chi},
     \quad\mbox{and}\quad
     \frac{\sqrt{|a_2|^2 + |b_2|^2}}{\Lambda^2}
\leq \frac{\pi}{m^2_\chi}\,,
\label{eq:naive_perturbativity_bound}
\end{align}
on the effective couplings versus the DM mass $m_\chi$.
The imposition of those constraints is a very loose statement
on the tree-level perturbativity.
If the limits are violated, we naively expect that
higher-loop corrections must be included,
which is beyond the scope of this paper.

\begin{figure}[ht!]
\centering
\includegraphics[scale=0.5]{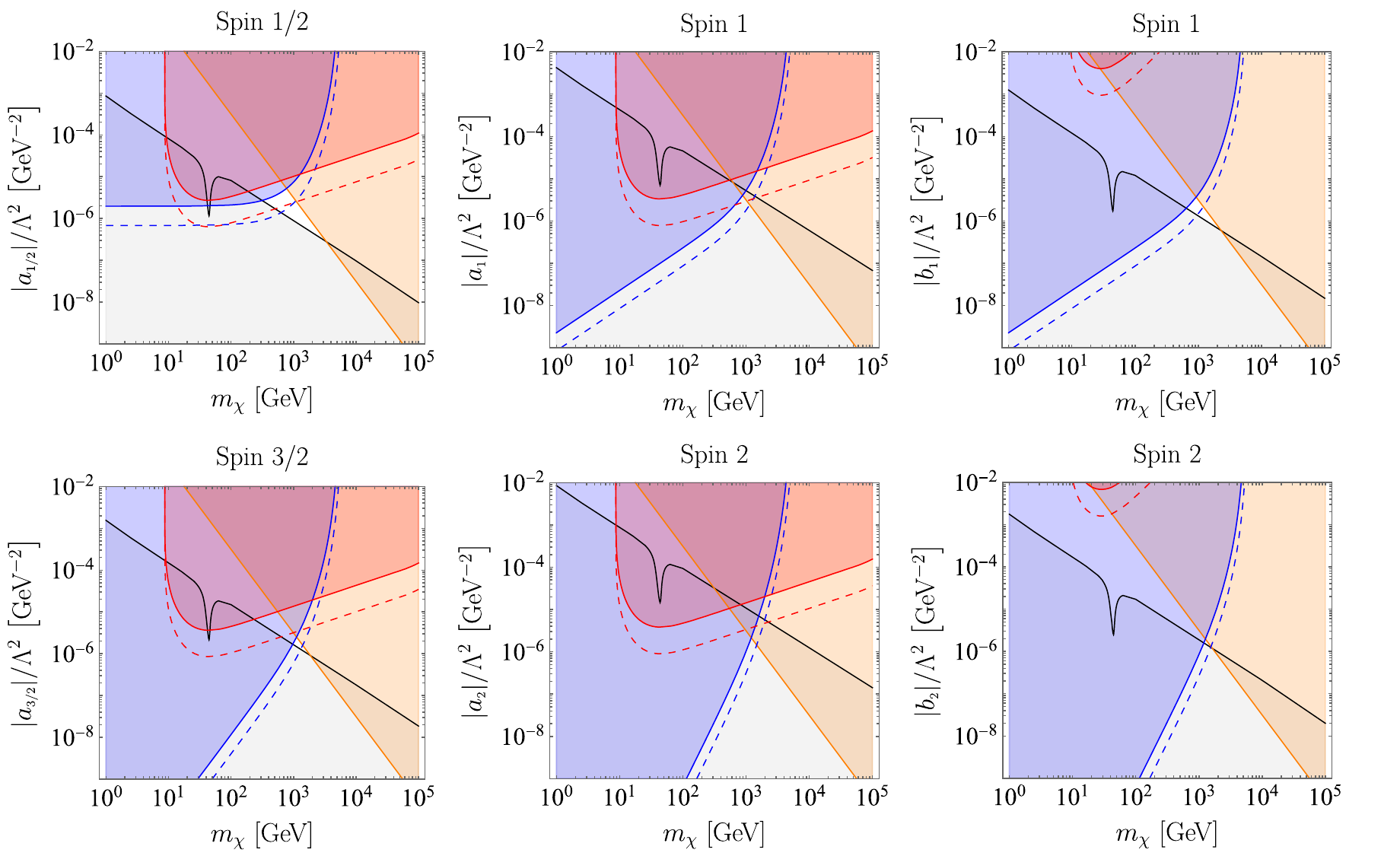}
\caption{\rm
    Combined exclusion limits on the effective couplings versus
    the DM mass $m_\chi$ from the measured DM relic abundance
    of the Planck satellite (black solid), the theoretical NPB
    (orange solid), the LHC (blue solid) and HL-LHC (blue dashed)
    mono-jet search experiments and the DM direct detection experiment XENONnT current limit
    with the 1.1 ton-year exposure
    (red solid) and the future prospect
    with the 20 ton-year exposure
    (red dashed). The top (bottom) left panel shows the
    combined constraint on the normalized coupling
    $|a_{1/2}|/\Lambda^{2}$
    ($|a_{3/2}|/\Lambda^{2}$)
    versus the DM mass in the spin-1/2
    (3/2) case. The top (bottom) middle panel
    shows the combined constraint on the normalized coupling
    $|a_1|/\Lambda^{2}$ ($|a_2|/\Lambda^{2}$) versus the DM mass,
    and the top (bottom) right panel
    shows the combined constraint on the normalized coupling
    $|b_1|/\Lambda^{2}$ ($|b_2|/\Lambda^{2}$) versus the DM mass
    in the spin-1 (2) case,
    while setting the other coupling to zero.
    }
\label{fig:combined_plots}
\end{figure}

Figure~\ref{fig:combined_plots} shows the combined exclusion limits
on the effective normalized couplings versus the anapole DM
mass $m_\chi$ from the Planck measurement of
the DM relic abundance (black solid), the theoretical NPB
condition (orange solid),
the mono-jet searches at
the LHC of 13 TeV with the integrated luminosity of 139 fb$^{-1}$
(blue solid) and the full running of the HL-LHC
of 14 TeV with the 3 ab$^{-1}$ integrated luminosity
(blue dashed),
and the present DM direct detection experiment XENONnT (red solid) with the 1.1 ton-year exposure
along with
the future XENONnT with the 20 ton-year exposure (red dashed).
The top (bottom) left panel is the combined constraint on
the normalized
coupling $|a_{1/2}|/\Lambda^{2}$ ($|a_{3/2}|/\Lambda^{2}$)
versus the mass $m_\chi$
in the spin-1/2 (3/2) case. The top (bottom) middle panel
shows the combined limit on the normalized coupling
$|a_1|/\Lambda^{2}$ ($|a_2|/\Lambda^{2}$)  and the top
(bottom) right panel shows the combined limit on
the normalized coupling
$|b_1|/\Lambda^{2}$ ($|b_2|/\Lambda^{2}$) versus
$m_\chi$ in the spin-1 (2) case.
Note that the future sensitivities of XENONnT can be reached
within about 5 years of running or the XLZD consortium of many
Xenon target experiment plans~\cite{Aalbers:2022dzr}.

In the spin-1/2 case, two regions are still allowed.
One tiny allowed region is near the $Z$ pole with $m_\chi=m_Z/2$,
where the relic abundance constraint becomes much weaker due to
the sharp $Z$-resonance effect. This tiny region is expected to be
completely excluded by the upcoming DM direct detection experiments
and the HL-LHC
experiment, as indicated by the red and blue dashed lines.
The other allowed region is a triangle-shape area centered near
$|a_{1/2}|/\Lambda^{2}=10^{-6}\, {\rm GeV}^{-2}$ and
$m_\chi = 1\, {\rm TeV}$, nearly the half of which can be probed
in the near future by XENONnT after the 20 ton$\cdot$yr exposure.
This shows the effectiveness of the complementary DM EFT approach.

In the spin-1 case, the $Z$-resonance region with $m_\chi\simeq m_Z/2$
is entirely ruled out by the combined constraints arising from both 
relic abundance measurements and LHC search experiments, illustrated in
the upper middle and right frames of Fig.$\,$\ref{fig:combined_plots}.
This complete exclusion arises from two synergistic effects, independent of the constraints posed by both the DM direct detection experiment and the NPB bound.
Compared to the spin-1/2 case, the relic abundance
constraint is bolstered by 1.5 times, attributed to a smaller spin-averaged
factor than that in the spin-1/2 case. Furthermore, the constraints imposed
by the LHC experiments are significantly heightened, primarily due to
the longitudinal mode of the spin-1 DM particle, particularly noticeable
for $m_\chi$ less than $1\, {\rm TeV}$.
Moreover, the spin-1 case with a non-zero $|a_1|$ but $|b_1|=0$
is expected to face total exclusion, as depicted in the top middle panel of
Fig.~\ref{fig:combined_plots}. Here, the combined constraints from the
relic abundance measurements, the (HL-)LHC and (upgraded) DM direct
detection experiments and the theoretical NPB synergistically contribute
to this complete exclusion. 
Conversely, the spin-1 case with a non-zero $|b_1|$ but $|a_1|=0$
features a triangular-shaped allowed region centered around
$|b_1|/\Lambda^{2}=10^{-6}\, {\rm GeV}^{-2}$ and $m_\chi = 1, {\rm TeV}$,
even though the permitted area is substantially reduced compared to
the spin-1/2 scenario, due to the much more strengthed constraints from
the relic abundance and LHC experiment.
While the complete HL-LHC operational phase
is not anticipated to entirely cover this small permissible region,
the future $100\, {\rm TeV}$ circular $pp$ collider~\cite{FCC:2018vvp}
promises the capability to thoroughly investigate and potentially close
off the remaining segments of this area, because of its vastly greater
collision energy and significantly improved sensitivity.

Despite the slightly weaker XENONnT constraints on the coupling
$|a_{3/2}|/\Lambda^2$ in the spin-3/2 case than the smaller spin cases
due to the spin-averaged and polarization-weighted 
factors, the currently allowed parameter region is smaller.
This is because the LHC constraint gets much stronger stemming
from the enhancement of the mono-jet production cross section
by the larger number of longitudinal modes as explained
in Sec.~\ref{sec:collider_searches}.

The full running of the HL-LHC is expected to probe nearly
half or more of the very tiny allowed regions.
This effect leads to a remarkable result for the higher spin
case, i.e., spin-2 DM.
As shown in the bottom-middle and bottom-right panels, the LHC
mono-jet searches so far have provided extremely strong
constraints: full exclusion for the $|a_2|/\Lambda^2$
dominated case and nearly exclusion except for the region
$m_\chi \sim 1.2$ TeV for the $|b_2|/\Lambda^2$ dominated case.
The remaining region is expected to be fully probed at the
HL-LHC in particular with more effective search strategies.
Due to the gradual tightening of constraints from relic
abundance and LHC data for higher-spin cases, it becomes evident
that all hypercharge anapole dark matter particles with masses
below the GeV scale are excluded, irrespective of their spin
values, provided they adhere to the thermal freeze-out scenario.

\begin{figure}[ht!]
\centering
\includegraphics[scale=0.60]{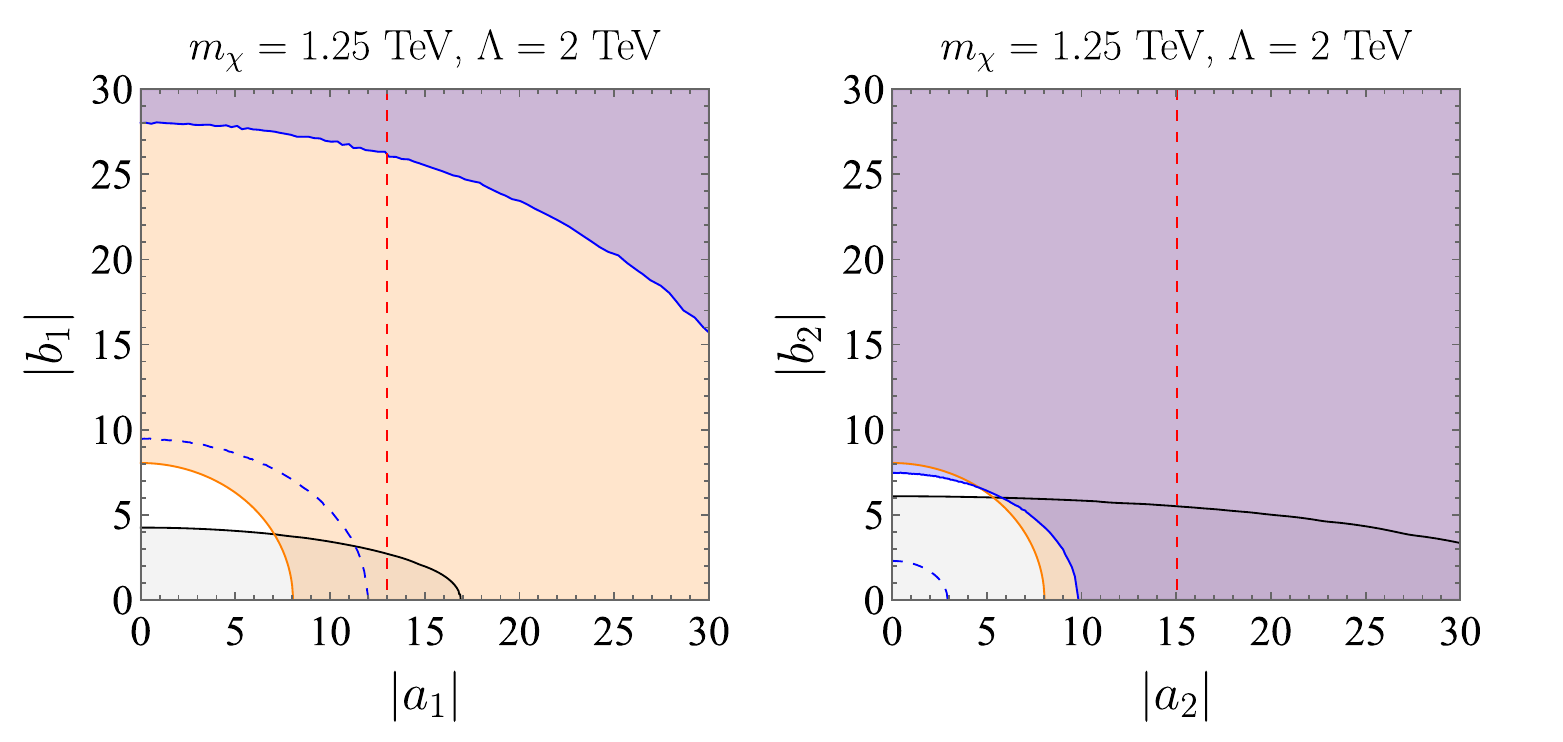}
\caption{\rm
    The parameter spaces of two couplings, $|a_1|$ and $|b_1|$ (left), and two couplings, $|a_2|$ and $|b_2|$ (right) for the specific values of the DM particle mass, $m_\chi=1.25\, {\rm TeV}$,
    and the cutoff scale, $\Lambda= 2\, {\rm TeV}$,
    in the spin-1 and 2 cases, respectively.
    The red and blue dashed lines indicate
    the projected sensitivities from the XENONnT
    with the 20 ton-year exposure
    and the HL-LHC experiment
    after the full running with the integrated luminosity of 3 ab$^{-1}$,
     respectively.
     }
\label{fig:contour_plots}
\end{figure}

In order to cover a general model set-up for spin-1 and 2 DM including both
non-negligible parity odd terms ($|a_{1,2}|/\Lambda^2$) and even terms
($|b_{1,2}|/\Lambda^2$) in the cross sections, we display the combined
constraints and sensitivities in the 2-dimensional ($|a_i|$ and $|b_i|)$
plane after fixing the DM mass $m_\chi = 1.25$ TeV and the cutoff scale
$\Lambda = 2$ TeV in Fig.~\ref{fig:contour_plots}. 
Each of the constraints is given by an ellipse due to
the cross sections are proportional to the combination of the absolute
squares of two couplings in the spin-1 and spin-2 cases.
We ignore the constraints on $|b_i|$ from XENONnT which are too weak.
The left panel is for the spin-1 and the right panel is for the spin-2 DM.
Note that the future XENONnT sensitivity with the 20 ton$\cdot$yr exposure
(red dashed vertical line) is comparable to those 
of the HL-LHC full running
for the spin-1 DM, while the latter becomes more powerful for the spin-2 DM. 
We expect that a higher spin $s > 2$ DM scenario would suffer from even 
stronger bounds although further dedicated studies are needed. 
The right panel of Fig.~\ref{fig:contour_plots} shows that the full running of 
the HL-LHC can probe fully the spin-2 scenario with $m_\chi=1.25$ TeV.

\section{Summary and conclusion}
\label{sec:summary_conclusion}

Our investigation focused on a scenario where DM is
characterized
as a Majorana particle possessing a non-zero spin, interacting
solely with SM particles via hypercharge anapole terms.
This scenario renders us to pursue the combined analysis
from various experimental results in a complementary way
via the EFT approach.
For the experimental/observational constraints and
sensitivities, we applied the Planck measurement of the DM
relic abundance, the direct
detection experiment XENONnT, and the LHC mono-jet searches
together with the expected sensitivities at the HL-LHC and the
future XENONnT.
Because of the straightforward calculations and strong
theoretical foundations, we conducted a focused numerical
analysis on DM spins $s=1/2$, 1, 3/2, and $2$, making
comparative analyses among them for the first time
within the realm of anapole DM studies.
A succinct summary of the anapole Dark Matter scenarios and
experimental searches analyzed in this paper is presented in
Tab.~\ref{tab:summary_of_works_on_anapole_dm}, juxtaposed with
previous literature for comparison.
The expectation for the higher spin DM scenarios will be briefly discussed
at the end of this section.
Considering the theoretically allowed range of the EFT approach
together with a grain of salt, We demonstrate that the hypercharge
anapole DM scenarios are currently on the verge of being discovered
or ruled out.

\begin{table}[htb]
\caption{\rm Comparison between our current research and various
         previous studies on EM anapole DM and hypercharge
         anapole DM conducted through three main experiments:
         relic abundance analysis,
         searches at the LHC, and/or direct detection
         experiments. The works are referenced with
         corresponding citation
         numbers provided in the references. In the case of
         spin-1 EM anapole
         DM, each cross mark (\ding{56}) denotes that
         the relic abundance and
         LHC search aspects have not been explored to date.
         On the other hand, the red check
         mark ({\color{red} \ding{52}})
         signifies the comprehensive quantitative investigation
         of all the clarified experimental aspects
         in the spin-1. 3/2 and 2 cases in
         addition to the spin-1/2 case, undertaken
         in our current work.
        }
\vskip 0.5cm
\centering
\begin{tabularx}{\linewidth}
    {|@{} c |*6{>{\centering\arraybackslash}X|}@{}}
\hline
\rb{-0.3ex}{\bf Scenario}
           & \multicolumn{2}{c|}{\bf \, EM anapole DM }
           & \multicolumn{4}{c|}{\bf Hypercharge anapole DM}
\\ \cline{2-7}
\rb{0.5ex}{\color{blue} [\,Spin\,]}
        &  {\color{blue}\large 1/2}
        &  {\color{blue}\large 1}
        &  {\color{blue}\large 1/2}
        &  {\color{blue}\large 1}
        &  {\color{blue}\large 3/2}
        &  {\color{blue}\large 2}
\\ \hline\hline
\,  \rb{0.5ex}{\color{blue} Relic abundance} \,
      &  \rb{0.5ex}{\small \cite{Ho:2012bg}\,
         \cite{Latimer:2017lwm}\, \cite{Bose:2023yll}}
      & \rb{0.2ex}{\Large \ding{56}}
      & \rb{0.5ex}{\small \cite{Arina:2020mxo}} &
       \multicolumn{3}{c|}{\Huge $\phantom{\checkmark}$}
\\ \cline{1-4}
 \rb{-0.2ex}{\color{blue} LHC search}
       & \rb{-0.2ex}{\small \cite{Gao:2013vfa}\, \cite{Alves:2017uls}\,
         \cite{Florez:2019tqr}}
       & \rb{-0.7ex}{\Large \ding{56}}
       & \rb{-0.2ex}{\small \cite{Arina:2020mxo}}
       & \multicolumn{3}{c|}{\rb{-1.7ex} 
       {\!\!\!\!\!\!
       This work\;\;\color{red}\LARGE {\ding{52}}}}
\\ \cline{1-4}
\rb{2.1ex}{\color{blue} Direct detection}
        & {\small \cite{Ho:2012bg}\, \cite{Geytenbeek:2016nfg}\,
          \cite{Alves:2017uls}\,
          \rb{0.3ex}{\cite{Kang:2018oej}\, \cite{Bose:2023yll}}}
        & \rb{2.0ex}{\small \cite{Hisano:2020qkq}\, \cite{Chu:2023zbo}}
        & \rb{2.0ex}{\small \cite{Arina:2020mxo}}
       & \multicolumn{3}{c|}{}
\\ \hline
    \end{tabularx}
\label{tab:summary_of_works_on_anapole_dm}
\end{table}

The main results of the present work can be summarized with the following key points:
\begin{itemize}
\item
As easily expected, the relic abundance imposes stronger constraints on
$d$-wave terms than the $p$-wave terms since larger couplings are required to
obtain the right relic abundance.
The coefficients of the $p$-wave terms decrease as the spin of DM due to the
spin-averaged and polarization-weighted
factors in the annihilation cross sections:
1/4 ($|a_{1/2}|$), 1/9 ($|b_{1}|$), 5/72 ($|a_{3/2}|$), and 1/20 ($|b_{2}|$).
This induces stronger constraints as the DM spin increases.
In each case, the relic abundance constraint in the $Z$-pole resonance 
region is about a factor of 10 times weaker than the others.
\item
The LHC and HL-LHC mono-jet searches play the most crucial role
in probing the higher-spin anapole DM lighter than about 1 TeV.
This is due to the kinematic factor associated with
the longitudinal modes of the DM particle.
The mono-jet searches play an essential role, especially for
probing the parity-even terms of the spin 1 or 2 DM, i.e.,
$|b_{1,2}|/\Lambda^2$, since their contributions to the DM
direct detection cross sections are suppressed by the recoil
energies.
\item The momentum transfer in the DM direct detection process
      is sufficiently small and hence the dominant contribution is through 
      the $t$-channel photon exchange.
      The constraint on the parity-odd couplings
      $|a_{1/2}|/\Lambda^2$, $|a_{1}|/\Lambda^2$, $|a_{3/2}|/\Lambda^2$, and
      $|a_{2}|/\Lambda^2$  are comparable to each other, with minor deviations
      due to the spin-averaged and
      polarization-weighted factors, 1/2, 1/3, 5/18, and 1/4, respectively.
      The cross section,
      when considering only the spin-1 (2)
      parity-even
      coupling $|b_1|/\Lambda^2$ ($|b_2|/\Lambda^2$), is suppressed
      in the
      non-relativistic limit by the
      small kinetic factor due to the
      CP selection rule.
      Consequently, as illustrated
      in the top (bottom) right panel
      of Fig.~\ref{fig:direct_plots}, the constraint on
      $|b_1|/\Lambda^2$ ($|b_2|/\Lambda^2$) is
      highly suppressed by the recoil energy as $m_T E_R / 2m_\chi^2$.
      It is noteworthy that the
      result from the near future XENONnT with the 20 ton-year exposure
      is expected to explore the allowed region of
      space that coincides with the HL-LHC expectations
      for the spin-1/2 case, as shown in the top left panel
      of Fig.~\ref{fig:combined_plots}.
      Remarkably, the upcoming XENONnT experiment has the
      potential to achieve the extended coverage within
      approximately 5 years, which is significantly sooner
      than the projected full operational timeline of the
      HL-LHC. Thus, the future DM direct detection
      experiments, such as the planned XLZD consortium
      involving multiple Xenon target experiments with
      more than 20 ton-year exposure, could eventually
      probe beyond the allowed region bounded by
      the HL-LHC expectation at a faster pace, 
      in the spin-1/2 case.
\item The NPB condition is equally and
     approximately applied to all the spin-1/2, 1, 3/2, and 2 couplings.
      The lower NPB bound is inversely
      proportional to the square of the DM mass, as presented in
      Eq.~\eqref{eq:naive_perturbativity_bound}. It is important
      to note that a breach of this conceptual NPB suggests
      a breakdown of the EFT framework, potentially
      resolvable by a more fundamental UV
      theory~\cite{Cabral-Rosetti:2014cpa,Cabral-Rosetti:2015cxa,
      Ibarra:2022nzm}. Hence, the NPB bounds
      need to be accepted cautiously.
\end{itemize}

Overall, the combined analysis shows that the hypercharge anapole
coupling of a higher spin DM is more stringently constrained or
expected to be probed sooner than that of a lower spin DM as evident in
Fig.~~\ref{fig:combined_plots}.
We expect the 100 TeV proton-proton circular collider experiment (FCC),
which is currently under R\&D would have a potential to completely
probe the whole remaining parameter space of the anapole DM scenarios
considered here.
Note that our analysis result combining the constraints from the
observed relic abundance and the mono-jet searches at the LHC can be
extended to a lighter DM mass down to $\mathcal O (10\,{\rm MeV})$ as
long as the DM relic is determined by the freeze-out mechanism,
providing a powerful exclusion bound already.

The DM particle with its spin larger than
2 is regarded to be innately a composite particle in order
to avoid various conceptual problems such as
unitarity issues. Nevertheless, if the scale of compositeness
is significantly high, we expect that
scenarios with spins greater than 2
could potentially be completely ruled
out because of the substantial kinematic factor linked to an
increased number of longitudinal modes of the DM particle
in the anapole vertices. although no conclusive comments can be 
made yet before dedicated studies.

\setcounter{equation}{0}

\section*{Acknowledgments}
\label{sec:acknowledgments}

This work is supported by the Basic
Research Laboratory Program of the National Research Foundation
of Korea (Grant No. NRF-2022R1A4A5030362 for SYC, DWK, and SS).
SYC is supported in part by the Basic Science Research Program of
Ministry of Education through the National Research Foundation of
Korea (Grant No. NRF-2022R1I1A3071226).
SS is supported in part by the National Research Foundation of Korea
(Grants No. NRF 2020R1I1A3072747).
JJ is supported by a KIAS Individual Grant
(QP090001) via the Quantum Universe Center
at Korea Institute for Advanced Study.
The hospitality of APCTP during the program ``Dark Matter as a Portal
to New Physics 2024" is kindly acknowledged.

\appendix

\setcounter{equation}{0}

\section{An algorithm for constructing the anapole vertices}
\label{appendix:anapole_vertex_any_spin}

This appendix is devoted to a compact description of
an efficient and systematic algorithm for constructing
any U(1) gauge-invariant anapole
vertex of a virtual gauge boson and two identical Majorana
particles of any spin $s$.

Let us begin by constructing the wave tensor of a particle of
non-zero mass $m$ and any non-zero spin $s$. For a non-zero
integer spin $s=n$, the wave function of an incoming massive boson
with momentum $k$ and helicity $\lambda$ is given by a wave tensor
defined as a product of $n$ spin-1 polarization vectors by
\begin{align}
   \epsilon^{\alpha_1\cdots \alpha_n}(k,\lambda)
=  \sqrt{\frac{2^s(s+\lambda)!(s-\lambda)!}{(2s)!}}
   \sum_{\{\tau_i\}=-1}^1 \delta_{\tau_1+\cdots +\tau_n,\,\lambda}\,
   \prod^s_{j=1}
   \frac{\epsilon^{\alpha_j}(k,\tau_j)}{\sqrt{2}^{|\tau_j|}},
\label{eq:bosonic_wave_tensor}
\end{align}
and, for a half-integer spin $s=n+1/2$, the wave function of
the massive fermion with momentum $k$ and helicity $\lambda$
is given by two types of wave tensors as
\begin{align}
   u^{\alpha_1\cdots \alpha_n}(k,\lambda)
&= \sum_{\tau=\pm 1/2}\, \sqrt{\frac{s+2\tau \lambda}{2s}}\,
   \epsilon^{\alpha_1\cdots \alpha_n}(k,\lambda-\tau)\,
    u(k,\tau)
    \,\,\quad \mbox{with} \quad |\lambda-\tau|\leq n,
\label{eq:u-spinor_tensor}
\\
   \bar{v}^{\alpha_1\cdots \alpha_n}(k,\lambda)
&= \sum_{\tau=\pm 1/2}\, \sqrt{\frac{s+2\tau \lambda}{2s}}\,
   \epsilon^{*\alpha_1\cdots \alpha_n}(k,\lambda-\tau)\,
   \bar{v}(k,\tau)
   \quad \mbox{with} \quad |\lambda-\tau|\leq n,
\label{eq:v-spinor_tensor}
\end{align}
with the helicity $\lambda$ varying from $-s$ to $s$.
The $u$ tensor is for an incoming fermion and
the $\bar{v}=v^\dagger \gamma^0$
tensor for an incoming anti-fermion.\s

The wave tensors have several characteristic features.
The bosonic wave tensors are totally symmetric, traceless
and divergence-free:
\begin{align}
    \varepsilon_{\mu\nu\alpha_i\alpha_j}
    \epsilon^{\alpha_1\cdots \alpha_i\cdots \alpha_j\cdots \alpha_s}
    (k,\lambda)
&= 0,
\label{eq:symmetric}
\\
   g_{\alpha_i\alpha_j}
   \epsilon^{\alpha_1\cdots \alpha_i\cdots \alpha_j\cdots \alpha_s}
   (k,\lambda)
&= 0,
\label{eq:traceless}
\\
   k_{\alpha_i}
   \epsilon^{\alpha_1\cdots \alpha_i\cdots \alpha_s}(k,\lambda)
&= 0,
\label{eq:divergence_free}
\end{align}
with $i,j=1,\cdots,n$, as indicated clearly
by Eq.$\,$(\ref{eq:bosonic_wave_tensor}),
and the fermionic wave tensors satisfy the fermionic version of the
divergence-free condition
\begin{align}
  \gamma_{\alpha_i}\, u^{\alpha_1\cdots \alpha_i\cdots \alpha_n}
= \gamma_{\beta_i}\, v^{\beta_1\cdots \beta_i\cdots \beta_n}=0.
\label{eq:fermionic_divergence_free}
\end{align}
These four properties of the wave tensors are very effective in
constructing the general three-point vertices
as well as the gauge-invariant ones as demonstrated
in a series of works~\cite{Behrends:1957rup,Auvil:1966eao,
Caudrey:1968vih,Scadron:1968zz,Chung:1997jn,Huang:2003ym}.

In the present work, we deal with the U(1) gauge-invariant anapole
$\chi\chi B$ vertex of two Majorana particles of any spin and a
virtual hypercharge gauge boson. The U(1) gauge
invariance allows us to treat the virtual gauge boson $B^*$ as
an on-shell spin-1 particle of a varying mass $m_*$,
as every term proportional to the four-momentum appearing
in the numerator of the propagator is effectively killed off.
The DM pair annihilation is related to the amplitude for
the $s$-channel annihilation process $\chi\chi\to B^*$ and the
mono-jet events at the LHC and HL-LHC involve the decay process
$B^*\to\chi\chi$ while the DM direct detection involves
the $t$-channel transition $\chi\to \chi B^*$.

As all the three processes with different event topologies are
related through crossing symmetry, it is sufficient to simply
derive the covariant three-point vertex for  the decay of
a spin-1 particle $B^*$ of mass $m_*$ into two particles,
$\chi_1$ and $\chi_2$, of identical spin $s$ and mass $m$
\begin{eqnarray}
     B^*(p,\sigma)
\to  \chi_1(k_1,\lambda_1) + \chi_2(k_2,\lambda_2),
\label{eq:b_to_xx_process}
\end{eqnarray}
where $p$ and $\sigma$ are the $B^*$ momentum and helicity and
$k_{1,2}$ and $\lambda_{1,2}$ are the momenta and helicities of
two $\chi$ particles. The corresponding helicity
amplitudes can be written in the Jacob-Wick helicity
formalism~\cite{Jacob:1959at,rose2013elementary} as
\begin{align}
   \mathcal{M}^{B^*\rightarrow \chi_1\chi_2}_{\sigma;\lambda_1,\lambda_2}(\theta,\phi)
&\, =\, \mathcal{C}^1_{\lambda_1,\lambda_2}\,
   d^1_{\sigma,\lambda_1-\lambda_2}(\theta)\,
   e^{i(\sigma-\lambda_1+\lambda_2)\phi}
   \quad \mbox{with} \quad |\lambda_1-\lambda_2|\leq 1
\\
&\, =\,
    \bar{\psi}_1^{\alpha_1\cdots \alpha_n}(k_1,\lambda_1)
    \;\Gamma^{(s)}_{\alpha_1\cdots \alpha_n,\beta_1\cdots \beta_n;\mu}(p,q)
    \;\psi_2^{\beta_1\cdots \beta_n}(k_2,\lambda_2)
    \;\epsilon^{\mu}(p,\sigma),
\label{eq:decay_helicity_amplitude}
\end{align}
with the Wigner $d$ function in Ref.$\,$\cite{rose2013elementary}
and  the integer $n=s$ or $s-1/2$ for the bosonic or fermionic
particle. In the bosonic case with $s=n$, the two spin-$s$
wave tensors are
\begin{align}
   \bar{\psi}_1^{\alpha_1\cdots \alpha_n}(k_1,\lambda_1)
&= \epsilon^{*\alpha_1\cdots \alpha_n}(k_1,\lambda_1),
\\
    \psi_2^{\beta_1\cdots \beta_n}(k_2,\lambda_2)
&=  \epsilon^{*\beta_1\cdots \beta_n}(k_2,\lambda_2),
\end{align}
which can be derived explicitly with the complex conjugation in
Eq.$\,$(\ref{eq:bosonic_wave_tensor}). In the fermionic case
with $s=n+1/2$ the two spin-$s$ wave tensors are
\begin{align}
   \bar{\psi}_1^{\alpha_1\cdots \alpha_n}(k_1,\lambda_1)
&= \bar{u}^{\alpha_1\cdots \alpha_n}(k_1,\lambda_1),
\\
    \psi_2^{\beta_1\cdots \beta_n}(k_2,\lambda_2)
&=  v^{\beta_1\cdots \beta_n}(k_2,\lambda_2),
\end{align}
which can be derived explicitly by use of Eqs.$\,$(\ref{eq:u-spinor_tensor})
and (\ref{eq:v-spinor_tensor}).

Taking the procedure described in detail
in Refs.$\,$\cite{Choi:2021ewa,Choi:2021qsb}, we can construct
the covariant three-point vertex systematically.
Before writing down them in a compact form, we
introduce two fermionic basic operators defined by
\begin{align}
   P^{\pm}
&= \frac{1}{2}(1\mp \kappa \gamma_5),
\\
   W^{\pm}_{\mu}
&= \frac{1}{2}
  (\pm \kappa \gamma_{\bot\mu}+\gamma_{\mu}\gamma_5),
\end{align}
where $\kappa=\sqrt{1-4m^2/m^2_*}$ and
$\gamma_{\bot\mu} = g_{\bot\mu\nu}\gamma^\nu$ with the
orthogonal tensor
$g_{\bot \mu\nu}= g_{\mu\nu}-\hat{p}_{\mu}\hat{p}_{\nu}+
\hat{q}_{\mu}\hat{q}_{\nu}$
and also four bosonic basic operators
\begin{align}
   S^{0}_{\alpha\beta}
&= \hat{p}_{\alpha}\hat{p}_{\beta},
\\
   S^{\pm}_{\alpha\beta}
&= \dfrac{1}{2}
   \big[g_{\bot\alpha\beta}\pm i \langle \alpha\beta \hat{p}\hat{q}\rangle\big],
\\
   V^{\pm}_{1\alpha\beta;\mu}
&= \dfrac{1}{2}\hat{p}_{\beta}
   \big[g_{\bot\alpha\mu}\pm i \langle \alpha\mu \hat{p} \hat{q}\rangle \big],
\\
   V^{\pm}_{2\alpha\beta;\mu}
&= \dfrac{1}{2}\hat{p}_{\alpha}
   \big[g_{\bot\beta\mu}\mp i \langle \beta\mu \hat{p} \hat{q}\rangle \big],
\end{align}
with the totally anti-symmetric tensor
$\langle \alpha\beta \hat{q}\hat{p}\rangle
=\varepsilon_{\alpha\beta\gamma
\delta}\,\hat{p}^{\gamma}\hat{q}^{\delta}$
with the convention $\varepsilon_{0123}=+1$.

The covariant fermionic three-point vertex
with the half-integer spin $s=n+1/2$ is given by
\begin{align}
   [\Gamma_F]
& = \displaystyle\sum_{\tau=-n}^{n}
    \bigg\{[\,\hat{q}\,]\,
    \Big( \theta(\tau)\, f^0_{\tau+1}\, [P^+]
        +\theta(-\tau)\, f^0_{\tau-1}\, [P^-]\Big)
        \nonumber \\
&   \qquad\qquad      + f^+_\tau\, [W^+] + f^-_\tau\, [W^-]
    \bigg\}\, [S^{\hat{\tau}}]^{|\tau|}[S^0]^{n-|\tau|},
\label{eq:general_gamma_f_vertex}
\end{align}
in terms of the fermionic $f^0$ and $f^\pm$ form factors, and
the covariant bosonic three-point vertices with the integer
spin $s=n$ by
\begin{align}
  [\Gamma_B]
= \displaystyle\sum_{\tau=-n}^n
   \bigg\{b^0_\tau\, [\,\hat{q}\,]\, [S^{\hat{\tau}}]
  +\theta(|\tau|-1)
    \Big(b^+_\tau [V_1^{\hat{\tau}}]
        +b^-_\tau [V_2^{\hat{\tau}}]\Big)
    \bigg\}\,
    [S^{\hat{\tau}}]^{|\tau|-1}[S^0]^{n-|\tau|}.
\label{eq:general_gamma_b_vertex}
\end{align}
in terms of the bosonic $b^0$ and $b^\pm$ form factors,
and the non-negative integer $n$ with $\hat{\tau}=\tau/|\tau|$
satisfying $\hat{\tau}=+1$ for $\tau=0$, the step function
$\theta(x)=1$ or 0 for $x\geq 0$ or $x<0$. respectively.

For two identical Majorana particles
($\chi_1=\chi_2$) coupled to the U(1) gauge boson,
the corresponding covariant three-point vertices $\Gamma_F$
and $\Gamma_B$ must satisfy the identical-particle (IP) relations
\begin{align}
   C\Gamma^\mu_{F\beta,\alpha}(p,-q)C^{-1}
&= \Gamma^\mu_{F \alpha,\beta}(p,q)
   \quad\mbox{ for fermions},
\label{eq:fermionic_ip_relation}
\\[5pt]
   \Gamma^\mu_{B \beta,\alpha}(p,-q)
&= \Gamma^\mu_{B \alpha,\beta}(p,q)
   \quad\mbox{ for bosons},
\label{eq:bosonic_ip_relation}
\end{align}
with the charge conjugation operator $C=i\gamma^2\gamma^0$
satisfying  $CC^\dagger=1$ and $C^\dagger=C^T=-C$.
These IP relations lead to the following relations
on the form factors:
\begin{align}
   f^0_{\tau\pm 1}
&= 0, \quad
   f^+_\tau = f^-_\tau,
\\
   b^0_\tau
&= 0, \quad \,\,
   b^+_\tau = b^-_\tau,
\end{align}
with $\tau=-n,\cdots, n$ for the non-negative integer $n$.
As a result, we end up with the sum $W^++W^-$ in the fermionic
case and the sum $V_1^{\pm}+V_2^{\pm}$ in the bosonic case.
Furthermore, imposing the U(1) gauge invariance condition
on the summed expression, we can obtain the modified fermionic
and bosonic vector operators:~\footnote{The term $\hat{q}_{\mu}
(\hat{q}\cdot \gamma)$ in $A_{\bot \mu}$ vanishes when
coupled to the fermionic wave tensors due to their on-shell
conditions.}
\begin{align}
   A_{\mu}
&= \gamma_{\bot \mu}\gamma_5,
\\
   V^{\pm}_{\mu}
&= \dfrac{1}{2} \big[\hat{p}_{\alpha}g_{\bot\alpha\mu}
+\hat{p}_{\beta}g_{\bot\alpha\mu}
\mp i \langle \alpha\beta \mu\hat{q}\rangle_{\bot} \big],
\end{align}
with the orthogonal gamma matrix
$\gamma_{\bot \mu}=g_{\bot \mu\nu}\gamma^{\nu}$ and
the angle-bracket notation of the Levi-Civita tensor
$\langle \alpha\beta\mu \hat{q}\rangle_{\bot}
=g^{\quad\nu}_{\bot \mu} \langle \alpha\beta\nu \hat{q} \rangle$
where the odd-parity operator
$\langle \alpha\beta \mu \hat{q}\rangle_{\bot}$ is
obtained through the following effective replacement:
\begin{align}
\hat{p}_{\beta}\langle \alpha\mu \hat{p}\hat{q}\rangle
-\hat{p}_{\alpha}\langle \beta\mu \hat{p}\hat{q}\rangle
&\overset{\scriptsize\mbox{eff}}{=}
-\langle \alpha\beta\mu\hat{q} \rangle,
\end{align}
which is guaranteed when the vertex operators are coupled to
the wave tensors of two on-shell Majorana particles of spin $s$.
For the detailed derivation, we refer to the works in
Ref.~\cite{Boudjema:1990st,Choi:2021ewa,Choi:2021qsb,Choi:2021szj,
Hagiwara:1986vm}).

By gathering all the basic and derived operators, we can readily formulate the covariant representation of the effective hypercharge anapole vertices for a pair of identical Majorana particles with arbitrary spin
\begin{align}
   [\Gamma_B]
&= \displaystyle\sum_{\tau=-n}^{n}
   \theta(|\tau|-1)\,
   b_{\tau}[V^{\hat{\tau}}][S^{\hat{\tau}}]^{|\tau|-1}
   [S^{0}]^{n-|\tau|}
\quad \mbox{for an integer spin $s=n$},
\\
   [\Gamma_F]
&= \displaystyle\sum_{\tau=-n}^{n}
    f_{\tau}\,
    [A\,][S^{\hat{\tau}}]^{|\tau|}[S^{0}]^{n-|\tau|}
\quad \mbox{for a half-integer spin $s=n+1/2$},
\end{align}
which can be rendered equivalent to
Eqs.$\,$\eqref{eq:general_fermionic_gamma}
and \eqref{eq:general_bosonic_gamma} after a proper dimensional
adjustment with the momentum-squareds, $p^2$ and $q^2$,
and the cutoff scale $\Lambda$. The proper adjustment can easily be
made by use of two effective identities
\begin{eqnarray}
       g_{\alpha\beta}
\,=\, S^+_{\alpha\beta} + S^-_{\alpha\beta}
      - \frac{(p^2-q^2)}{q^2} \, S^0_{\alpha\beta}
      \qquad\mbox{and}\qquad
       p_{\alpha}p_{\beta}
\,=\, p^2 S^0_{\alpha\beta},
\end{eqnarray}
which are effectively valid due to their contractions with the wave tensors
of two on-shell Majorana particles.
As can be checked easily, the total number
of independent terms is $2s$ both for the fermionic and bosonic
cases as mentioned in Ref.~\cite{Boudjema:1990st}.

\section{Calculation strategy of the DM relic abundance}
\label{appendix:relic_calculation_strategy}

In this appendix, we provide a detailed
explanation of how to calculate the DM relic abundance.
This abundance is determined by the thermal freeze-out
processes illustrated in
Fig.~\ref{fig:diagrams_annihilations}.
In the absence
of phase transitions, the entropy of the Universe in a co-moving
frame, $S\equiv s a^3=(2\pi^2/45)g_s T^3 a^3$,
with the degrees of freedom $g_{s}$ contributing to the entropy
density $s$ is conserved throughout its evolution with the
decreasing temperature $T$ and, accordingly, the scale factor $a$.
The freeze-out occurs during the radiation-dominated epoch,
enabling us to
write down the energy density $\rho=(\pi^2/30)g_\rho T^4$, with
the degrees of freedom $g_{\rho}$, involved in the Hubble parameter
$H=(1/a)\, da/dt=\sqrt{8\pi G\rho/3}$  with the gravitational
constant $G$. In this case, the yield $Y(x)$ of DM particles,
varying over $x=m_\chi/T$ (that is, over time and/or temperature),
satisfies the so-called evolution equation:
\begin{align}
   \frac{dY}{dx}
= -\frac{\lambda_{\rm ann}}{x^2}\frac{g_s}{\sqrt{g_\rho}}
   \bigg[1+\frac{1}{3}\frac{d(\ln g_s)}{d(\ln T)}\bigg](Y^2-Y_{eq}^2),
\label{eq:y_evolution_equation}
\end{align}
where $\lambda_{\rm ann}=(\pi/45)^{1/2}\, M_{pl}\, m_\chi \langle\,
\sigma v_{\scriptsize\mbox{M\o l}} \rangle$ with
the so-called M\o ller velocity $v_{\scriptsize\mbox{M\o l}}$
of two annihilating DM particles~\cite{Gondolo:1990dk} , and
$Y=n/s$ and $Y_{eq}=n_{eq}/s$ including
the DM number density $n$ and the thermal-equilibrium density
$n_{eq}$  in the co-moving frame. Here, the yield $Y_{eq}$ of
the DM particles in thermal equilibrium can be written as
\begin{align}
  Y_{\scriptsize\mbox{eq}}(x)
= 0.1154\frac{g_\chi}{g_s}x^2K_2(x),
\label{eq:y_thermal_equilibrium}
\end{align}
with the DM spin degrees of freedom $g_\chi$ and the second-kind
modified Bessel function $K_2$ of order 2.
The simple asymptotic
expression at large values of $x \gg 1$ is
\begin{align}
        Y_{\scriptsize\mbox{eq}}(x)
\approx 0.145\frac{g_\chi}{g_s}x^{3/2}e^{-x}.
\label{eq:y_asymptotic_expression}
\end{align}

In the present work, we calculate the thermally-averaged cross
section $\langle \sigma v_{\scriptsize\mbox{M\o l}} \rangle$ by
adopting its explicit form~\cite{Gondolo:1990dk} as
\begin{align}
\langle \sigma v_{\scriptsize\mbox{M\o l}} \rangle
&=\frac{x}{8 m^5_{\chi} K_2^2(x)}
\int^{\infty}_{4m^2} ds \sqrt{s}(s-4m^2_{\chi})K_1(\sqrt{s}/T)\, \sigma(s),
\end{align}
involving the annihilation cross sections in
Eqs.~\eqref{eq:spin-1/2_annihilation_xsection}
or~\eqref{eq:spin-1_annihilation_xsection} with
the second-kind modified Bessel function $K_1$ of order 1.
This is because the conventional series expansions of the cross
sections over the relative velocity between two annihilating
DM particles do not converge when the annihilation energy
$\sqrt{s}$ is close to the $Z$ boson mass due to the sharp
$Z$-boson pole contribution~\cite{Griest:1990kh}.
The exponentially-stiff suppression
of the modified Bessel functions for large values of $x \gg 1$
renders the numerical calculation unreliable. Instead,
we use its analytic asymptotic expression directly as
\begin{align}
\langle \sigma v_{\scriptsize\mbox{M\o l}} \rangle
&\sim \frac{x^{3/2}}{8 m^5_{\chi} }\sqrt{\frac{2m_{\chi}}{\pi}}
\int^{\infty}_{4m^2_{\chi}} ds \;s^{1/4}(s-4m^2_{\chi})\,
\mbox{exp}\bigg[x\bigg(2-\frac{\sqrt{s}}{m_{\chi}}\bigg)\bigg]
\Bigg(1-\frac{15}{4x}+\frac{2m_{\chi}}{8x\sqrt{s}}\Bigg)
 \sigma(s),
\label{eq:averaged_x_section_asymptotic_expression}
\end{align}
for $x \gg 1$.
The exclusion limits on the couplings satisfying the observed
DM relic density
$\Omega_\chi h^2\approx 0.12$~\cite{Planck:2018vyg} can be derived
by integrating out the equation \eqref{eq:y_evolution_equation}
from $x=1$ to $x=10^3$ with the initial condition
$Y=Y_{eq}$ at $x=1$. Numerically, the function $\lambda_{\rm ann}$
dependent on the annihilation cross section varies stiffly with
the DM mass $m_\chi$ in the GeV region, reducing the accuracy of
the numerical calculation~\cite{Steigman:2012nb}.
The numerical accuracy can be enhanced greatly by taking the
integration after recasting the evolution equation as
\begin{align}
   \frac{dW}{dx}
= -\frac{\lambda_{\rm ann}}{x^2}
   \frac{g_s}{\sqrt{g_\rho}}
   \bigg[1+\frac{1}{3}\frac{d\ln g_s}{d\ln T}\bigg]
   \big[e^W-e^{(2W_{eq}-W)}\big],
\end{align}
with a more slowly varying logarithmic function $W=\ln Y$.
The relic abundance calculations described so far are
obtained semi-analytically, and the results are consistent
with the \texttt{MicrOMEGAs}~\cite{Belanger:2018ccd}
calculations.

\bibliography{ref}

\end{document}